\documentclass[smallcondensed, nospthms]{svjour3}

\usepackage{amsmath}
\usepackage{amssymb}

\allowdisplaybreaks

\usepackage{mathtools}

\usepackage{amsthm}

\usepackage{graphicx}
\usepackage{epstopdf}
\usepackage{subfigure}
\usepackage{multirow}

\usepackage{natbib}
\bibpunct{(}{)}{;}{a}{}{,}

\usepackage{appendix}
\usepackage[table, x11names]{xcolor}
\usepackage{longtable}
\usepackage{tabu}
\usepackage{color}
\usepackage[justification=centering]{caption}
\usepackage{umoline}

\journalname{Celestial Mechanics and Dynamical Astronomy}

\title{Dynamical Lifetime Survey of Geostationary Transfer Orbits}
\titlerunning{Dynamical Survey of GTOs}

\author{Despoina K. Skoulidou \and
			Aaron J. Rosengren \and \\ 
			Kleomenis Tsiganis \and 
			George Voyatzis}
\authorrunning{D. K. Skoulidou et al.}

\institute{D. K. Skoulidou \and K Tsiganis \and  G. Voyatzis
\at Department of Physics, Aristotle University of Thessaloniki, 54124 Thessaloniki, Greece  \\ 
\email{dskoulid@physics.auth.gr}
\and A. J. Rosengren
\at Aerospace and Mechanical Engineering, University of Arizona, Tucson, AZ 85721, USA}

\date{Received: date / Accepted: date}

\begin{document}
\maketitle

\begin{abstract}
In this paper, we study the long-term dynamical evolution of highly-elliptical orbits (HEOs) in the medium-Earth orbit (MEO) region 
around the Earth. The real population consists primarily of Geosynchronous Transfer Orbits (GTOs), launched at specific inclinations, 
Molniya-type satellites and related debris. We performed a suite of long-term numerical integrations (up to 200~years) within a realistic dynamical model, aimed primarily at recording the dynamical lifetime of such orbits (defined as the time needed for atmospheric 
reentry) and understanding its dependence on initial conditions and other parameters, such as the area-to-mass ratio ($A/m$). Our results 
are presented in the form of 2-D lifetime maps, for different values of inclination, $A/m$, and drag coefficient. We find that the majority 
of small debris ($>70\%$, depending on the inclination) can naturally reenter within 25-90~years, but these numbers are significantly less optimistic for large debris (e.g.,\ upper stages), with the notable exception of those launched from high latitude (Baikonur). We estimate 
the reentry probability and mean dynamical lifetime for different classes of GTOs and we find that both quantities depend primarily and strongly on initial perigee altitude. Atmospheric drag and higher $A/m$ values extend the reentry zones, especially at low inclinations. 
For high inclinations, this dependence is weakened, as the primary mechanisms leading to reentry are overlapping lunisolar resonances. 
This study forms part of the EC-funded (H2020) ``ReDSHIFT'' project.

\keywords{Satellites \and Geosynchronous Transfer Orbits \and Disposal orbits \and Dynamical evolution and stability}
\end{abstract}

\section{Introduction}
\label{intro}

Explorer VI (1959 Delta 1), with an apogee of 48 700 km, perigee of 6649 km, and equatorial inclination of $47^\circ$, represented the first satellite to be deliberately launched into a highly eccentric orbit \citep{dK87}, prompting the first detailed studies of the perturbing gravitational forces of the Sun and Moon acting on near-Earth satellites. The effects of these distant ``third bodies'' were often assumed negligible in comparison to the perturbing forces due to the Earth's oblateness, but in the case of highly elliptical orbits (HEOs), lunisolar perturbations were found to change the elements of the orbit to a large degree over extended periods of time. Kozai had found that lunisolar perturbations shortened the orbital lifetime of Explorer VI by a factor of ten, and Musen et al. showed that it could be as little as a month, depending on the time of day of launch \citep{pM59}. \\

Thus, the Sun and the Moon may provide a substantial perigee boost for the satellite under properly chosen circumstances. For other conditions, depending on the particular phase angles \citep[cf.][]{pM59}, the perturbation may be minimized to obtain a relatively stable orbit. Explorer XII (1961 Upsilon) was the first satellite to be launched at a time that was preselected for a minimum orbital lifetime of a year, utilizing lunisolar perturbations to ensure relative stability \citep{rSbS62}. The lifetime of a satellite in a highly eccentric orbit therefore depends critically on the initial conditions \citep{pM59, gCdS67, dK75, gJeR76}, and realistic lifetime predictions must generally consider gravitational perturbations, atmospheric drag, and solar radiation pressure as acting simultaneously throughout the orbital evolution; this implies that the area-to-mass ratio ($A/m$) can be critical.\\

There exists already a vast literature on lifetime estimation dating to before the Sputnik era \citep{jS57, bSjC66, gCdS67, dK78, dK82, dK88}. Lifetime predictions of satellite orbits primarily disturbed by drag (i.e., in LEO region) can be derived, e.g., from simplified analytical formula that determine the rate of change of the orbital period and assume a slow lowering of perigee into the denser regions of the atmosphere. For highly eccentric and distant satellites, however, it is rather necessary to resort to numerical integrations. It has been noticed from such studies that the dynamical lifetime does not have an obvious relation to the initial orbital elements, as the interactions between the gravitational perturbations on these orbits are complex \citep{gJ91, kS95, rS04, vM12, yWpG16, yWpG17}. The perturbing force of Earth's oblateness changes the orientation of an orbit with respect to the disturbing planes of the Moon and Sun, thereby modifying the phase of the lunisolar perturbation on the eccentricity. The orbit will be terminated by atmosphere drag, when the perigee has lowered into dense regions of the atmosphere. Note that the regime of the solar activity may change a lot during the predicted lifetime, as the typical values of the thermospheric density change one order of magnitude between minimum and maximum of solar activity,  let alone strong solar events. And such variations are really very difficult to handle in the propagations of the equations of motion, except using  numerical integration with a solar activity realistic enough through usual proxies such as F10.7cm.  However, a precise predictive study of GTOs evolution is not the scope of this work. Rather, a statistical comparison between different inclination bands is sought for.\\

Past numerical simulations have shown that some geostationary transfer orbits (GTOs) are dynamically short-lived \citep{gJ91,kS95,vM12}, and recent works have demonstrated that this problem still presents a number of interesting and hitherto unknown features \citep{yWpG16, yWpG17}.  Motivated by our preceding dynamical survey of orbital stability and associated lifetimes in circumterrestrial space -- from low-Earth orbit (LEO) to beyond the geostationary ring (GEO) \citep{aR18} -- conducted in the framework of the ReDSHIFT project \citep{ReDSHIFT,ReDSHIFT18}, we revisit the problem of GTO long-term stability. In fact, we extend our previous results, based solely on the long-term conservative evolution, by adding the effect of atmospheric drag at high altitudes. \\ 

We consider the evolution of highly-elliptical orbits (HEOs) in the medium-Earth orbital zone. The real population consists primarily of GTOs (launched from different sites), Molniya satellites and related debris. Hence, we focus on fictitious objects started at four different inclination values, corresponding to the latitudes of the main launching sites, and we study their long-term (200 years) dynamical evolution. We then display dynamical maps of orbital lifetimes, and study their characteristics as we vary physical parameters. Previous studies have investigated deorbiting through the use of solar radiation pressure (SRP) perturbations \citep[e.g.,][]{jAdT06, cL13}, generally focusing on direct removal, requiring relatively high area-to-mass ratios. Here, we study whether de-orbiting can be achieved through the deployment of a realistically sized solar sail at the end of the nominal mission, and the long-term coupling between non-gravitational forces and gravitational resonances \citep{jrD16, yWpG17}. The existence of a (fixed) sail is modeled here by a suitably enhanced $A/m$ ratio. Finally, we present statistics on the mean dynamical lifetime and probability of reentry, for the different families of objects. The set-up of our simulations is discussed in Section 2, while our results are presented in Section 3. Our conclusions are summarized in Section 4.

\section{Model and Numerical set-up}
\label{sec:1}

\subsection{Geostationary transfer orbit environment}
\label{subsec:21}

Geostationary transfer orbits are able to cross both the LEO and GEO regions, having semi-major axes roughly between $20100$~km and $28000$~km (or $a\in\left(0.498,0.664\right)\ a_{GEO}$) and eccentricities between $0.5$ and $0.8$. Their inclinations are representative of their launch-site latitude; e.g., the ESA station at Kourou ($i\sim5.235^{\circ}$), KSC at Cape Canaveral ($i\sim28.533^{\circ}$), and Baikonur Cosmodrome ($i\sim46^{\circ}$). With similar orbital characteristics as the GTOs, the critically inclined Molniya satellites in medium-Earth orbits (MEOs) are located at the critical inclination $i_c\sim 63.4^{\circ}$, with orbital periods close to the 2:1 commensurability with the Earth's rotation rate. \\

\begin{figure}[htp!]
	\captionsetup{justification=justified}
	\centering
	\includegraphics[width=6.0cm,height=4.75cm]{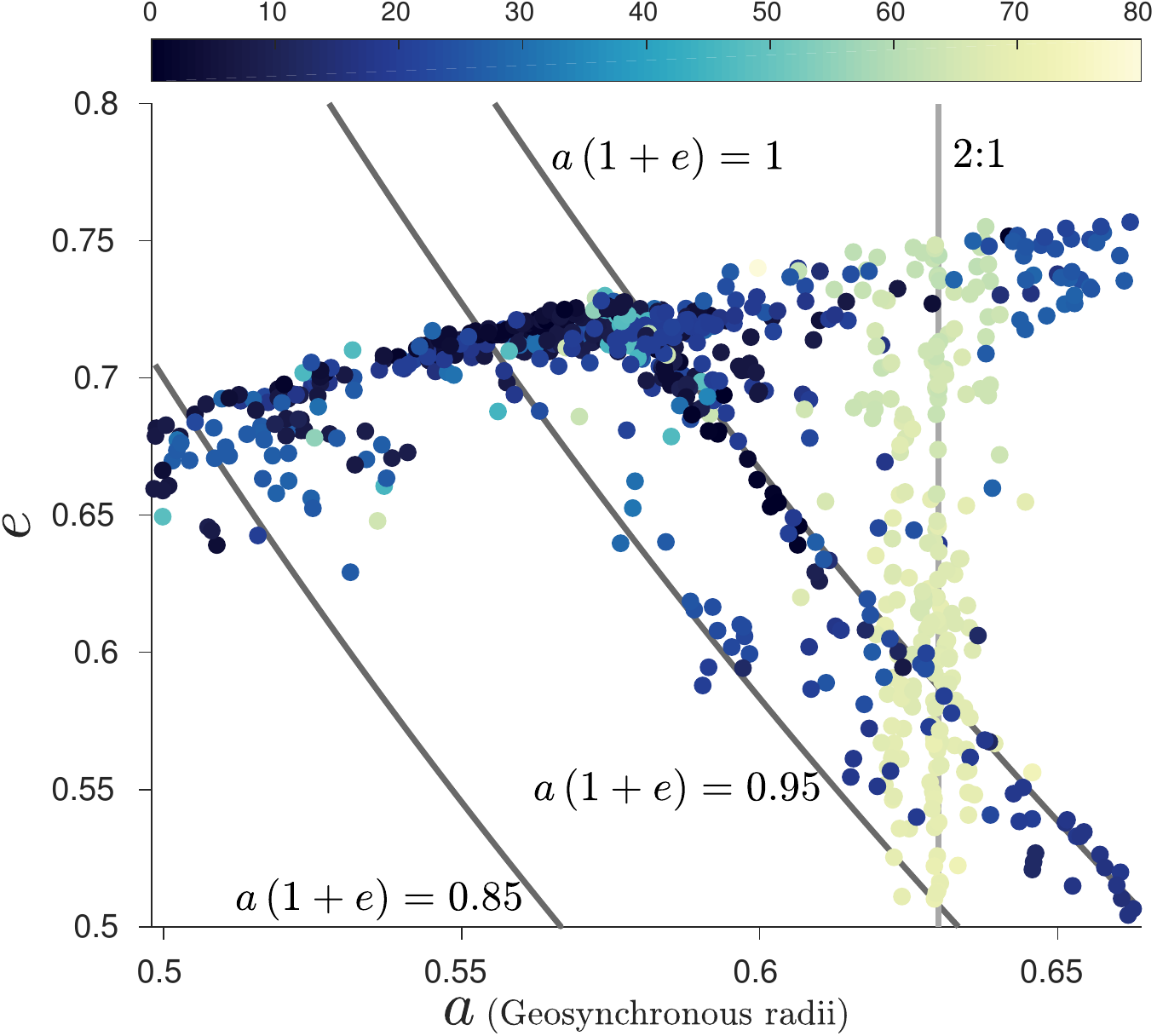}
	\includegraphics[width=6.0cm,height=4.75cm]{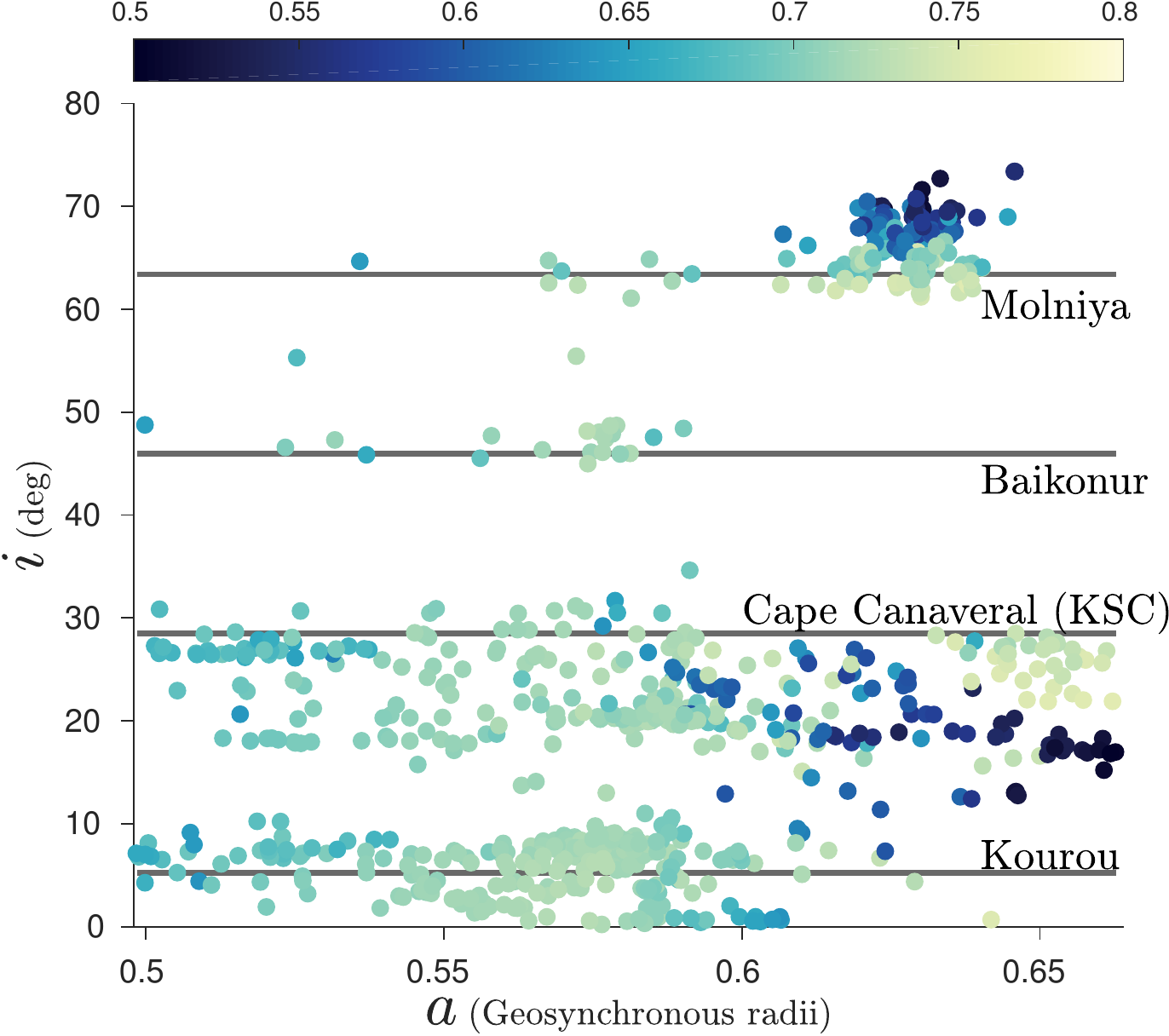}
	\caption{The cataloged resident space objects in the semi-major axis--eccentricity and semi-major axis--inclination space, where the colorbar corresponds to the missing `action' element in each two-dimensional plot (dark blue corresponds to low values, while yellow indicates high values). The gray vertical curve corresponds to $a_{2:1res}$. The darker gray curves in the $a-e$ plot correspond to apogee values of $0.85$, $0.95$, and $1\ a_{GEO}$, respectively. The horizontal gray curves in the $a-i$ plot correspond to inclination values near the respective latitudes of Kourou ($\sim5.235^{\circ}$), Cape Canaveral ($\sim28.533^{\circ}$), and Baikonur ($\sim46^{\circ}$), and Molniya's critical value ($\sim63.4^{\circ}$). (Resident Space Object Catalog. www.space-track.org. Assessed 25 Oct. 2016.)}
	\label{fig:norad_inc_ecc}      
\end{figure}

\begin{figure}[htp!]
	\captionsetup{justification=justified}
	\centering
	\includegraphics[width=6.0cm,height=4.75cm]{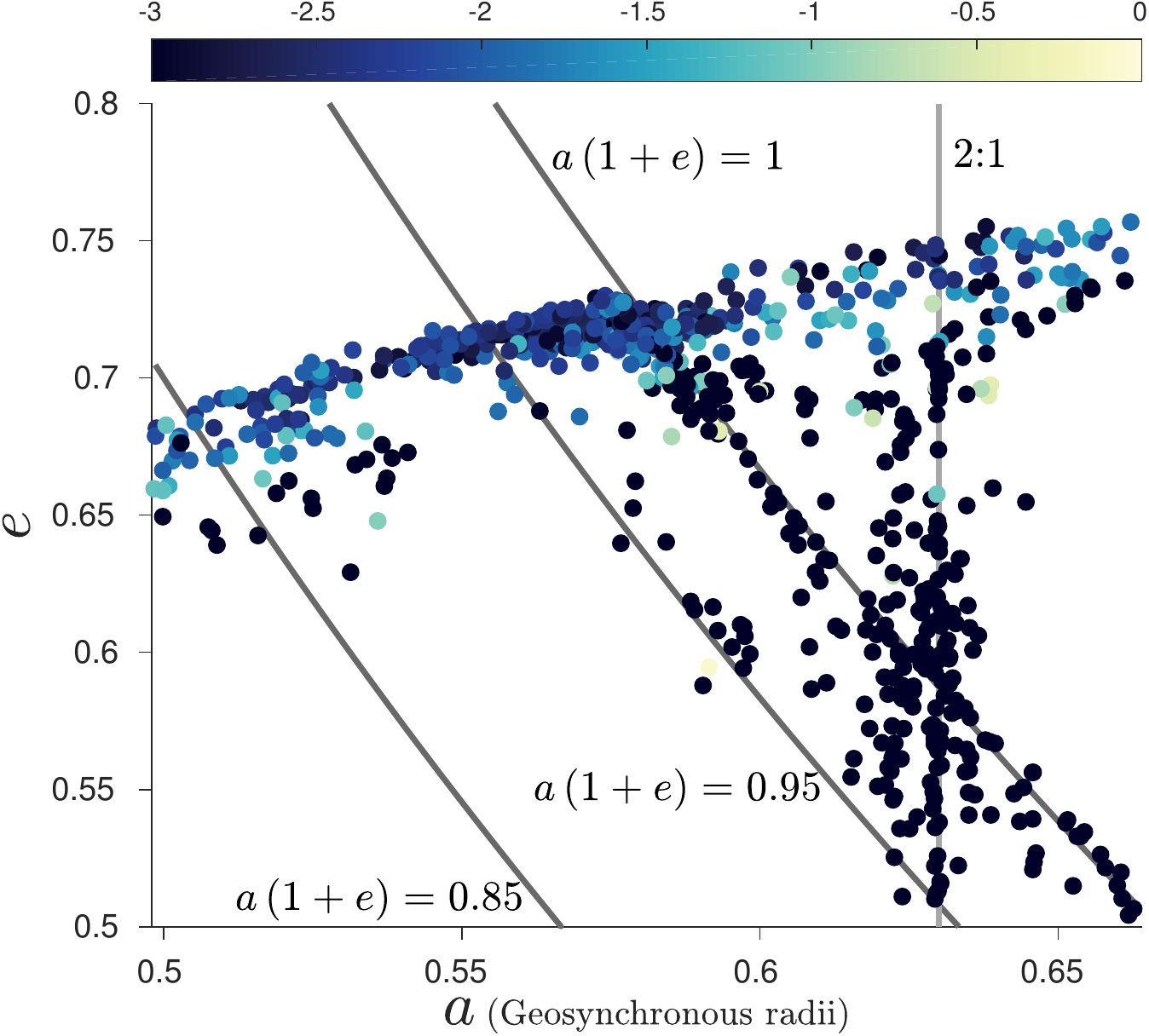}
	\includegraphics[width=6.0cm,height=4.75cm]{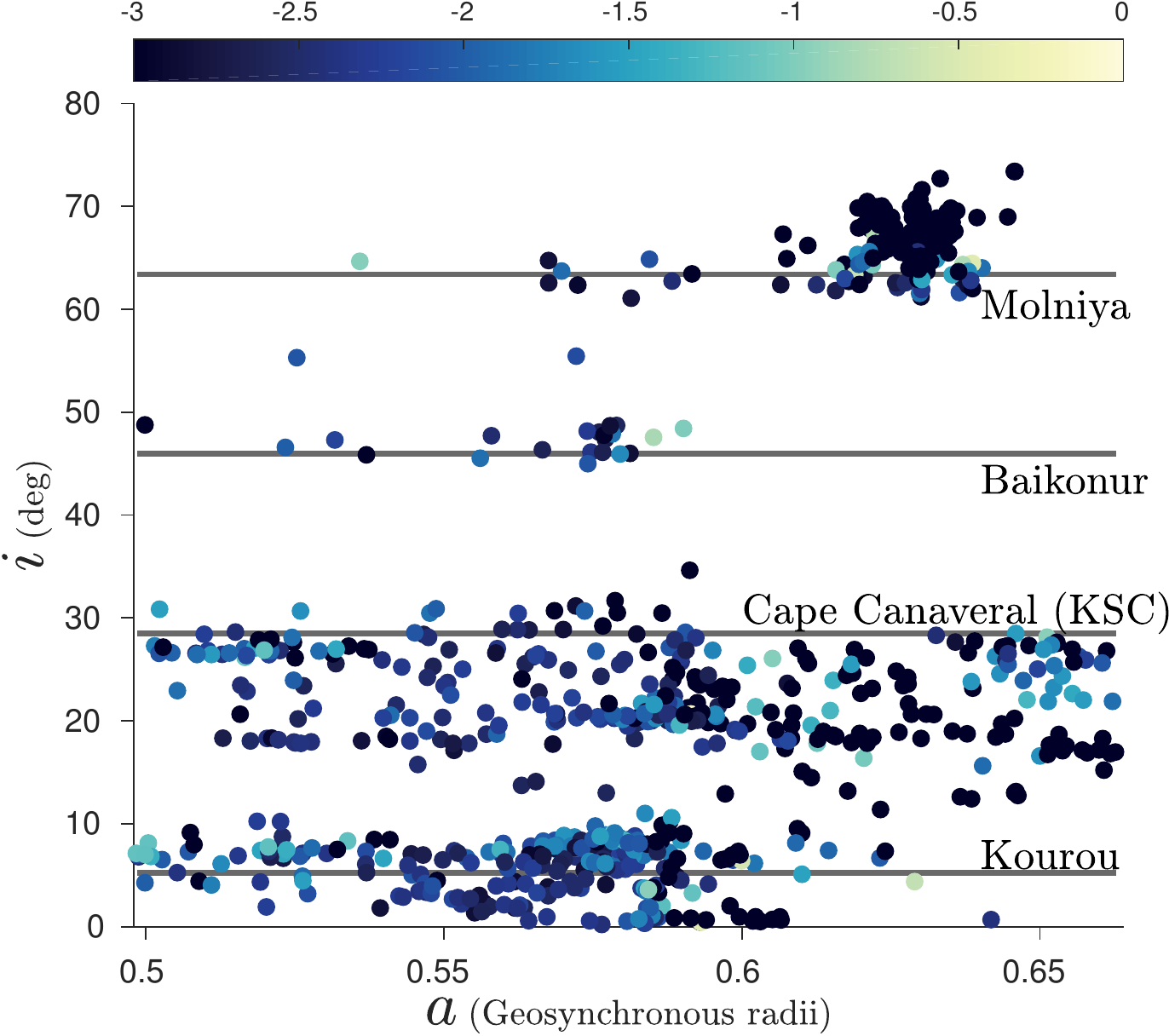}
	\caption{The cataloged resident space objects in the semi-major axis--eccentricity and semi-major axis--inclination space, where the colorbar corresponds to logarithm of $A/m\in\left[0:1\right]$m$^2$/kg (dark blue corresponds to low values, while yellow indicates high values). Please note that $\log\left(A/m\rightarrow 0\right)\rightarrow\infty$, hence, for presentation purposes, we set a minimum value of -3 in the colorbar. The darker gray curves in the $a-e$ plot correspond to apogee values of $0.85$, $0.95$ and $1\ a_{GEO}$. The horizontal gray curves in  the $a-i$ plot correspond to inclination values near the latitudes of Kourou ($\sim5.235^{\circ}$), Cape Canaveral ($\sim28.533^{\circ}$), and Baikonur ($\sim46^{\circ}$) and Molniya's critical inclination ($\sim63.4^{\circ}$). (Resident Space Object Catalog. www.space-track.org. Assessed 25 Oct. 2016.) }
	\label{fig:norad_amr}      
\end{figure}

The Resident Space Object Catalog\footnote{www.space-track.org. Assessed in 25 Oct. 2016., provided by JSpOC (Joint Space Operations Center)},  provides information on orbital and physical characteristics of Earth's satellites and space debris with sizes larger than $10$ cm. Figure~\ref{fig:norad_inc_ecc} shows a snapshot of the cataloged objects, focused on  the GTO region, projected in $a-e$ (left) and $a-i$ (right) space, where the colorbar denotes inclination and eccentricity, respectively. The Resident Space Object Catalog follows the \textit{Two-line element} format (i.e., TLE). According to \citet{dV13}, the TLE set provides the $B^{*}$ coefficient, which is related  to the true ballistic coefficient, 
\begin{equation}\label{eq:1}
BC =\frac{m}{C_{d}A}, 
\end{equation}
through equation 
\begin{equation}
BC=\frac{R_{E}\rho_{o}}{2~B^{*}},
\end{equation}
where $R_{E}=6378.135$~km is Earth radius, and $\rho_{o}$ is the atmospheric density at perigee (assumed to be $2.461\cdot 10^{-5}~kg/m^2/ER$ with $ER=6375.135$~km), $A$ is the cross-sectional area of the object, $m$ is its mass and $C_{d}$ is the 
drag coefficient. Then, the effective area-to-mass ratio, $C_R \left(A/m\right)$, can be computed by eq.\ref{eq:1}, assuming here $C_{d}=2.2$ and $C_{R}=1$ for reflectivity coefficient. Figure~\ref{fig:norad_amr} shows the same elements-space plots, but with the colorbar instead denoting the logarithm of the effective area-to-mass ratio, $C_R \left(A/m\right)$ (up to $1$ m$^2$/kg). Hereafter, we will use `$A/m$' instead of `$C_R \left(A/m\right)$' when we will refer to the effective area-to-mass ratio. \\

Resident space objects are divided into three main sub-populations, according to their $A/m$ value. Payloads and upper stage launch vehicles typically have $A/m$ of $0.02$ m$^2$/kg, while other debris have $\sim 1$ m$^2$/kg, and high area-to-mass ratio (HAMR) objects \citep[q.v.,][]{rM08} are in excess of $1$ m$^2$/kg. There are currently 842 objects registered by the Resident Space Object Catalog in the region $0.498\le a \le 0.664\ r_{GEO}$ and $0.5\le e \le 0.8$, three of which can be classified as HAMRs (i.e., $A/m> 1$ m$^2$/kg) and have been excluded from our study. Among 839 objects with $A/m\in\left[0,1\right]$ m$^2$/kg, 742 ($88.5\%$) have $A/m\le 0.02$ m$^2$/kg, and 97 ($11.5\%$) have $A/m\in\left(\left. 0.02,1\right]\right.$ m$^2$/kg. \\

Table~\ref{tab:objects} shows the number of cataloged objects, sorted by inclination in blocks of size $\left[i_{lat}-5^{\circ},i_{lat}+5^{\circ}\right]$, where $i_{lat}$ is inclination of the launch-site of Kourou, KSC, and Baikonur, or the critical value of Molniya orbits, and $A/m$ values in range $0-0.02$ or $0.02-1$ m$^2$/kg. In section \ref{sec:2}, we compare our results with the distribution of cataloged objects shown in Table \ref{tab:objects}.\\

\begin{table}[htp!]
	\captionsetup{justification=justified}
	\centering
	\caption{Number of cataloged objects, sorted according to their inclination and $A/m$ values.}
	\label{tab:objects}
	\begin{tabular}{cccc}
	\hline\noalign{\smallskip}
	latitude/inclination & $i_{lat}$ & $N\left(A/m\le0.02\right)$ 
		& $N\left(A/m\in\left(\left.0.02,1\right.\right]\right)$ \\
	Kourou & $5.235^{\circ}$ & 227 & 40  \\
	KSC & $28.533^{\circ}$ & 130 & 19 \\ 
	Baikonur & $46^{\circ}$ & 18 &  3 \\ 
	Molniya & $63.4^{\circ}$ & 129 & 16 \\
	\noalign{\smallskip}\hline
	\end{tabular}
\end{table}

Many objects have apogee close to $1\ a_{GEO}$, as expected for GTOs, and low $A/m$. Note that, as we have verified by a short numerical simulation, GTOs with apogee equal to $a_{GEO}$ can evolve to lower apogees ($\sim 0.95\ a_{GEO}$, or less, at high inclination), as reflected in the 
cataloged objects. At the same time, the inclination would vary within a range of $\sim 5^{\circ}$ over $\sim20$~years.\\

\subsection{Dynamical model and grid definition for numerical study}
\label{subsec:22}

\begin{figure}[htp!]
 \centering
 \includegraphics[width=0.7\textwidth]{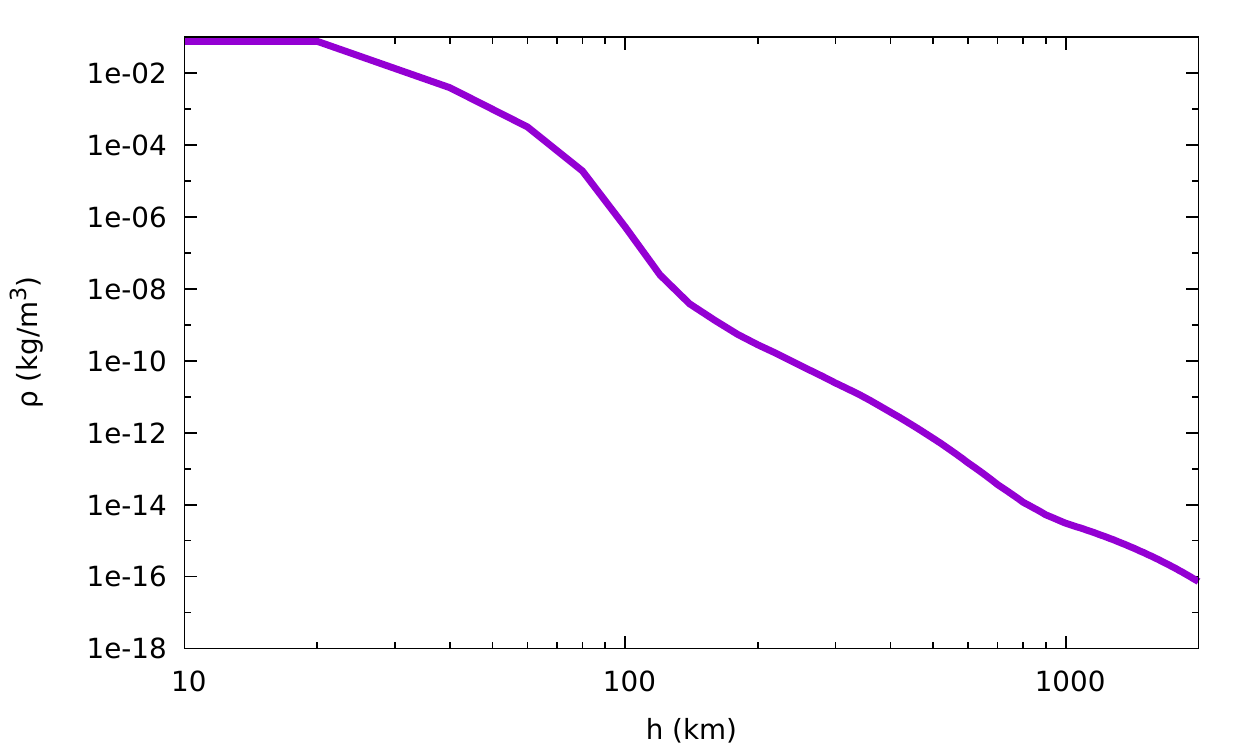}
 \caption{Atmospheric density, $\rho$, as function of altitude, $h$, according to the model adopted here.}
 \label{fig:density}
\end{figure}

In our study, we use a dynamical model that accounts for the gravitational potential of Earth up to degree and order 2 (i.e., $J_{20}$, $J_{22}$), the Moon and Sun as perturbing bodies, and direct solar radiation pressure, assuming the ``cannonball model''. We did not include shadow effects, as they are negligible far away from the LEO region. In LEO, however, atmospheric drag plays a significant role in the evolution of these low-altitude satellites. Since GTOs are designed to reach both LEO and GEO, we performed simulations that included atmospheric drag in our dynamical model, in an effort to assess its effect on the distribution of dynamical lifetimes. To this purpose, we decided to use a basic exponential atmospheric drag model, as described in \citet{dV13}, where the atmospheric density depends on the altitude only, and the data  are taken from the US Standard Atmosphere (1976) and CIRA-72 models\footnote{see Table 8-4 in \citet{dV13}}, where a moderate solar activity is assumed. In Fig.~\ref{fig:density}, the atmospheric density, $\rho$, as function of altitude, $h$, is shown, where it is clear that the density decays exponentially with increasing altitude and falls off drastically beyond a few hundred km; however, on centennial time-scales, we expect the phenomenon to be non-negligible. We also assumed the drag coefficient to be constant and equal to $2.2$ (i.e., $C_{d}=2.2$).\\

For our numerical integrations, we use our SWIFT-SAT symplectic integrator, which is based on the mixed-variable symplectic integrator of \citet{jWmH91}, as included in the SWIFT package of \citet{hLmD94}. SWIFT-SAT uses the non-averaged equations of motion and is suitable for dynamical studies of bodies with negligible mass, orbiting an oblate central body and perturbed by other massive bodies. Though originally designed for modeling heliocentric motions, SWIFT has been adapted in the past for dynamical studies of natural satellite systems \citep{dN03,aM12}. This requires the Sun to be treated as a massive, distant satellite of the central planet, in order to account for its gravitational perturbations, and a rotation and translation of the coordinate system, to align the fundamental plane to the planet’s equator and to set its origin to the center of the planet. In SWIFT-SAT, the positions of the Moon and the Sun are given as an ephemerides, numerically produced in a precise integration of the entire planetary system, as described in \citet{dS17} and \citet{aR18} in order to avoid mismatches that may occur during a direct numerical evolution of the test particles with Earth (as opposed to the massive Sun) as central body. We have additionally modified SWIFT to properly account for the dynamics in the  circumterrestrial environment by incorporating the Earth’s ellipticity perturbations ($J_{22}$) and solar radiation pressure (cannonball model without shadowing effects). We refer the reader to \citet{dS17} and \citet{aR18} for a more in-depth discussion of SWIFT-SAT and particular validations that were performed to ensure effective performance. In addition, SWIFT is able to incorporate weakly dissipative effects, such as was done by \citet{wB01} for the Yarkovsky effect for asteroids dynamics, or atmospheric drag for satellite dynamics (as used in SWIFT-SAT). The code is no longer symplectic of course, but it behaves smoothly, if the non-conservative kick is added symmetrically before the first and after the last sub-step of the symplectic algorithm.\\

Our grid of initial conditions is shown in Table~\ref{tab:init_cond}. Initial inclinations where chosen to represent the launch-site latitudes for GTOs or the critical inclination of Molniya orbits. We chose a more wider grid in ($a,e$) than the distribution of cataloged GTOs, as presented in \ref{subsec:21}, in order to investigate the dependence of lifetime, as well as to get a more global view of HEO dynamics. As the initial orbit orientation angles ($\Omega$ and $\omega$) also affect secular evolution, we chose $16$ different configurations\footnote{Note that the satellite's initial node and perigee angles were taken relative to the ecliptic lunar values at each epoch.}.  Finally, the initial mean anomaly was set to $M=0$. The runs were performed twice, for two preselected epochs (JD 2458475.2433, denoted hereafter as ``Epoch 2018'', and JD 2459021.78, hereafter ``Epoch 2020'') and for two different values of $A/m$; a typical one for spent upper stages (0.02 m$^2$/kg) and an augmented one (1 m$^2$/kg), representing a satellite equipped with a large sail, or small debris.\\

We performed all runs both with and without atmospheric drag in the model (hereafter denoted as \textit{DRAG} or \textit{NO DRAG} cases, respectively). All integrations spanned 200 years, using a time step $dt=0.004$~days (sidereal), with the reentry limit set to $R_{E}+100$~km. All numerical simulations were performed in a cluster of PCs provided by AUTH Computer Infrastructure and Resources. Hence, in total, a set of $\sim$ 2 million orbits were propagated, amounting to an equivalent of $\sim$4.5 years of CPU time.\\

\begin{table}[htp!]
	\captionsetup{justification=justified}
	\centering
\caption{Grids of initial conditions for the GTO study using SWIFT-SAT, 
	for dynamical maps in $a - e$ phase space.}
\label{tab:init_cond}
 \begin{tabular}{ll}
\hline\noalign{\smallskip}
 $a$ $(a_{GEO})$ & $0.498 - 0.664$  \\
      $\Delta a$ & $0.00475$        \\ 
             $e$ & $0.5 - 0.8$      \\ 
      $\Delta e$ & $0.015$          \\ 
       $i$ $(^\circ)$ &  $3.235 - 7.235$ \\
 \multirow{ 3}{*}{ }  & $26.533 - 29.533$ \\ 
                      & $44 - 48$ \\ 
                      & $61.4 - 65.4$ \\  
$\Delta i$ $(^\circ)$ & $1$ \\
$\Delta \Omega$ ($^\circ$) & $\left\{ 0, 90, 180, 270 \right\}$ \\
$\Delta \omega$ ($^\circ$) & $\left\{ 0, 90, 180, 270 \right\}$ \\ 
$C_R (A/m)$ (m$^2$/kg) & $\left\{ 0.02, 1 \right\}$ \\
\noalign{\smallskip}\hline
 \end{tabular}
\end{table}

\section{Results}
\label{sec:2}

\subsection{Dynamical lifetime maps}
\label{subsec:31}

\begin{figure}[htp!]
	\captionsetup{justification=justified}
	\centering 
	\begin{tabular}{cc}
	\textbf{NO DRAG} & \textbf{DRAG}\\
  \multicolumn{2}{c} {$\bf i_{o} = 5.235^\circ$}\\
  \includegraphics[width=6.0cm,height=4.75cm]{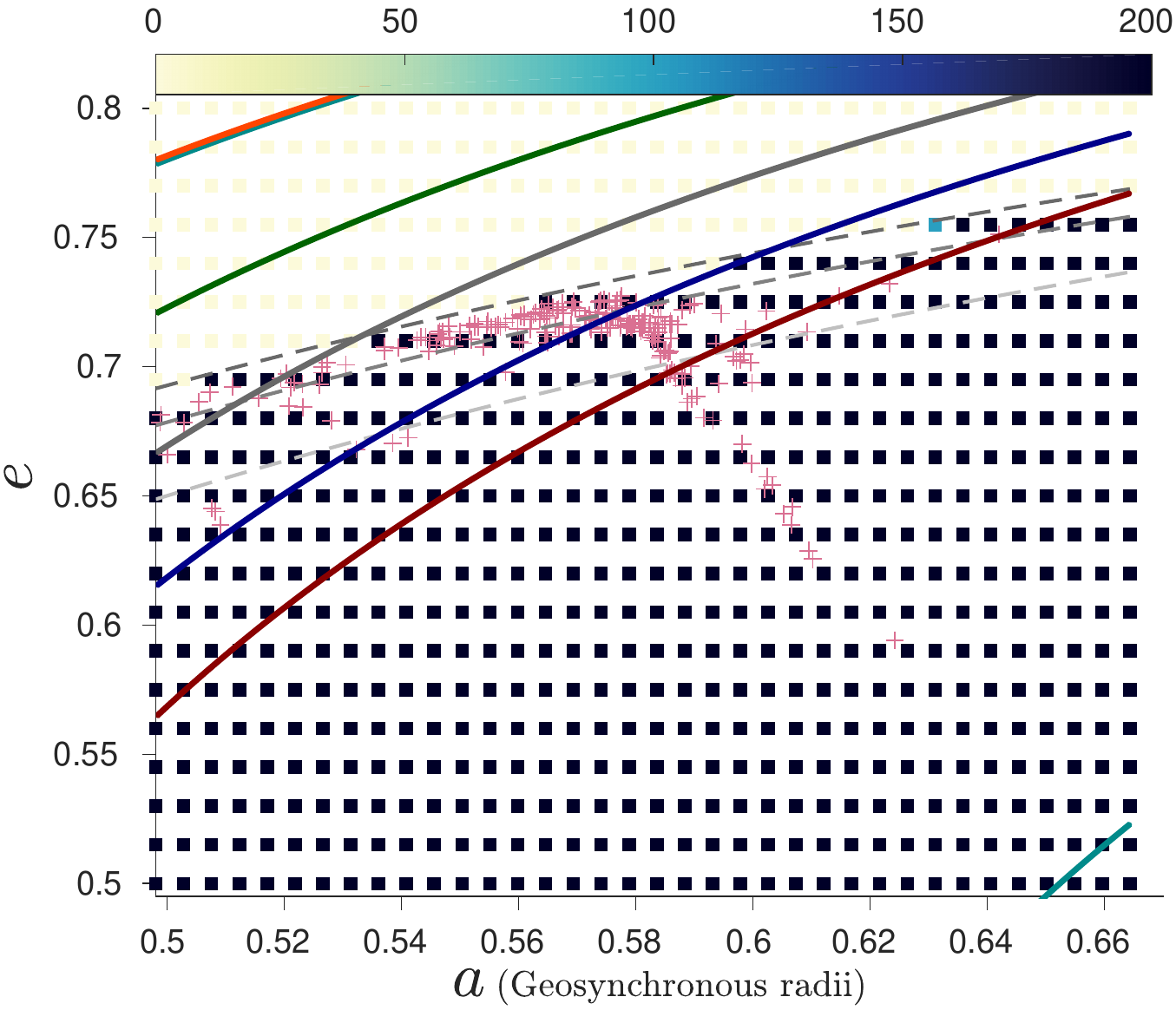} &
  \includegraphics[width=6.0cm,height=4.75cm]{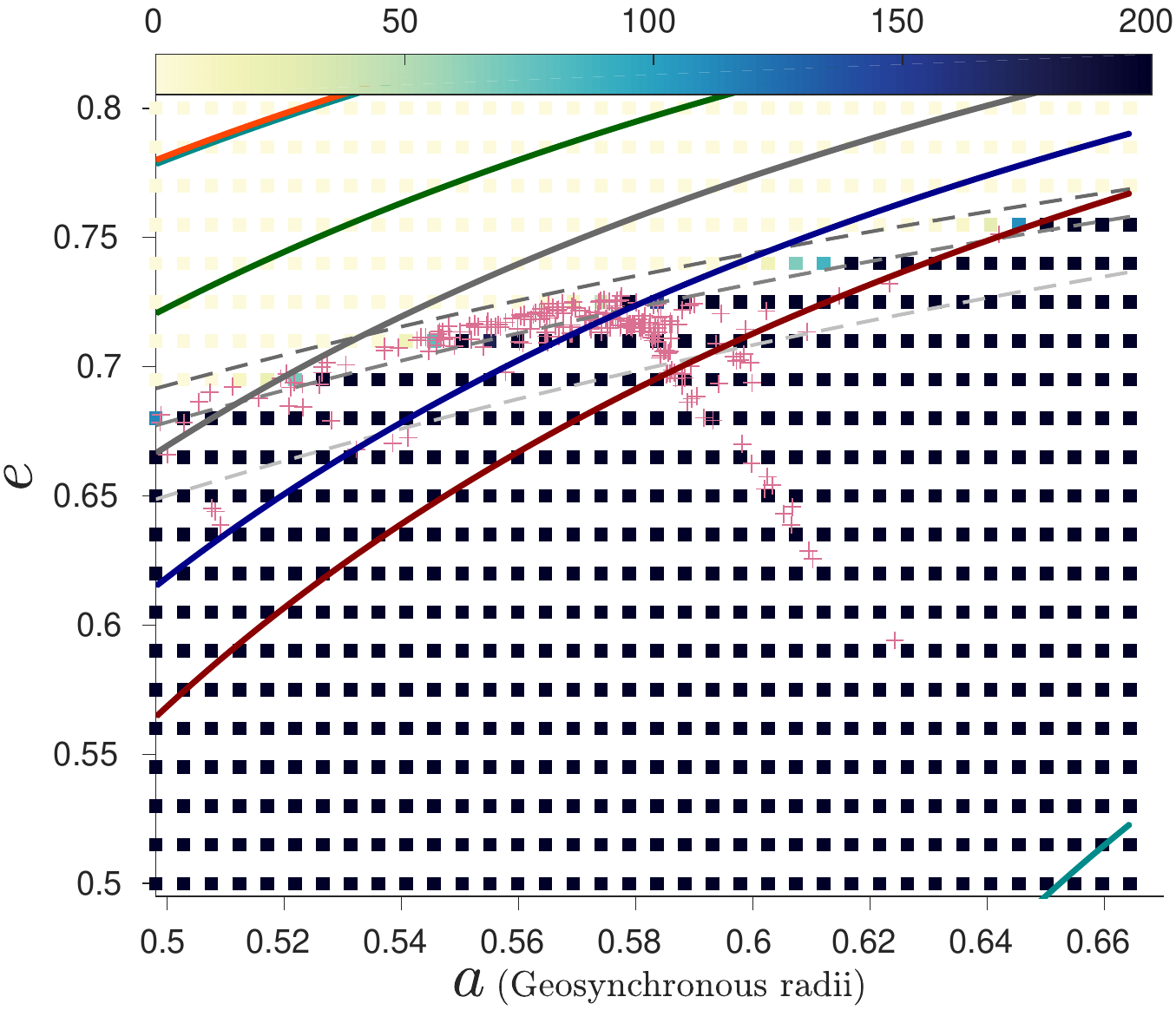} \\
  \multicolumn{2}{c} {$\bf i_{o} = 28.533^\circ$}\\
  \includegraphics[width=6.0cm,height=4.75cm]{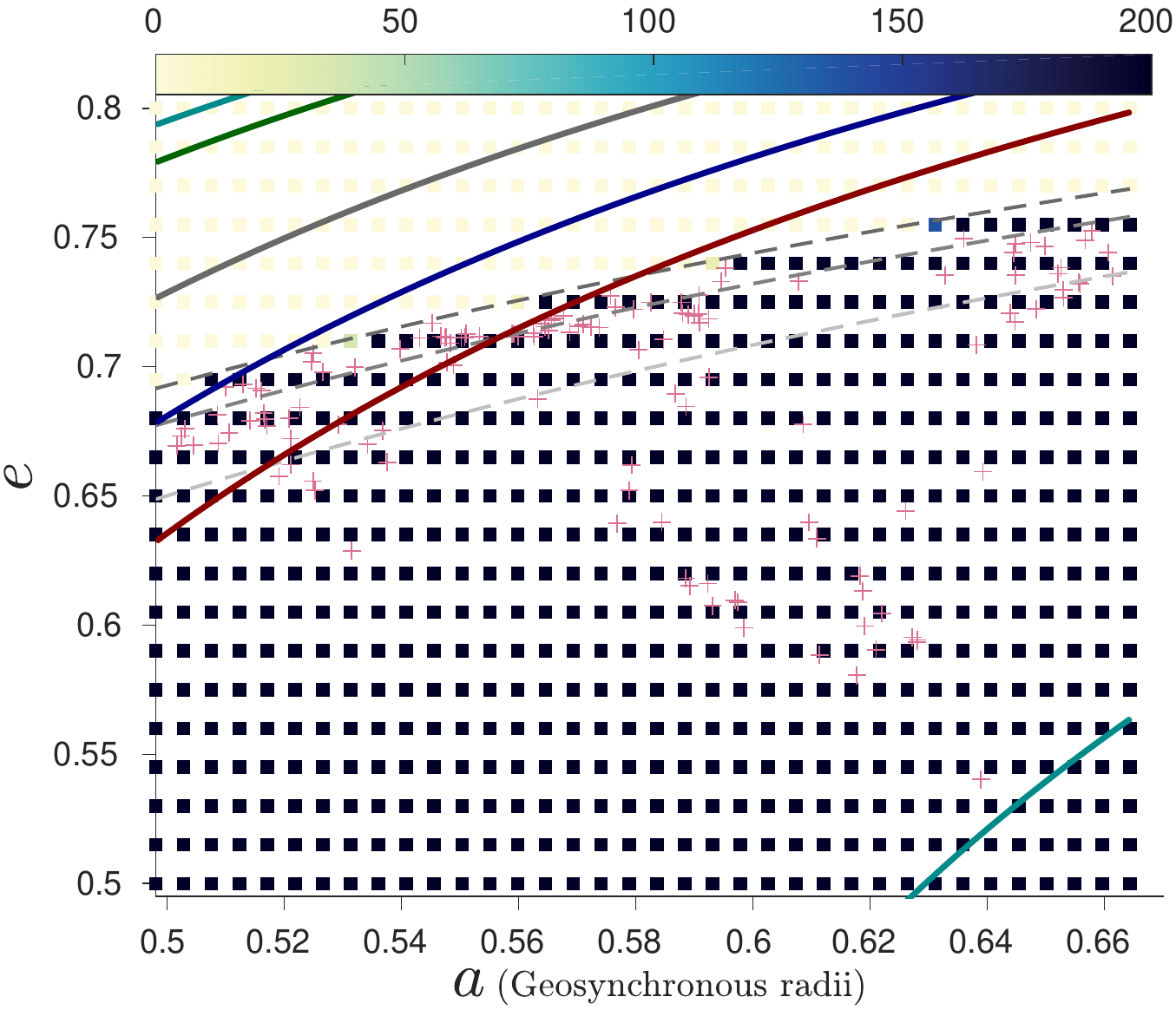} &
  \includegraphics[width=6.0cm,height=4.75cm]{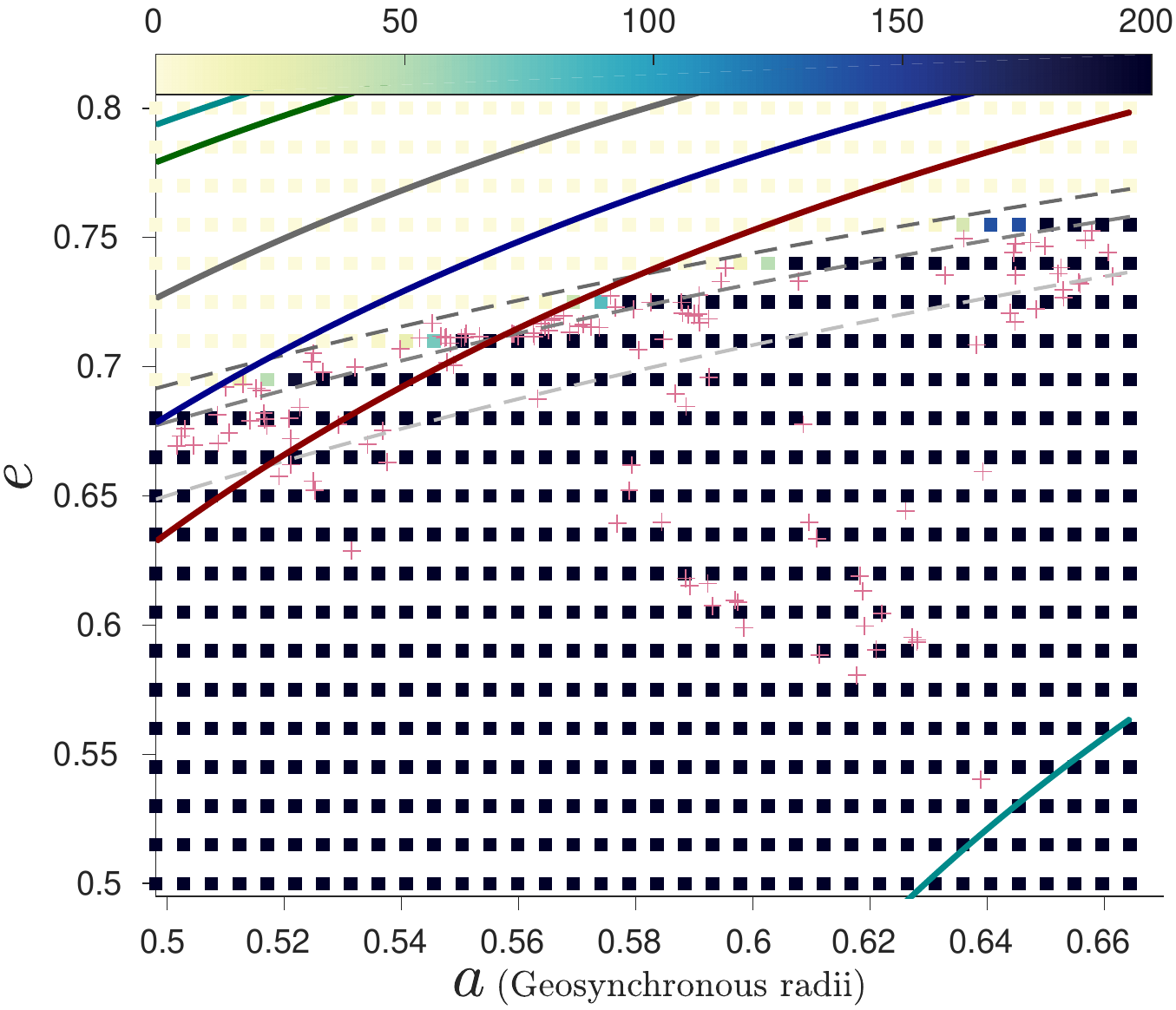}\\
  \multicolumn{2}{c} {$\bf i_{o} = 46^\circ$}\\
  \includegraphics[width=6.0cm,height=4.75cm]{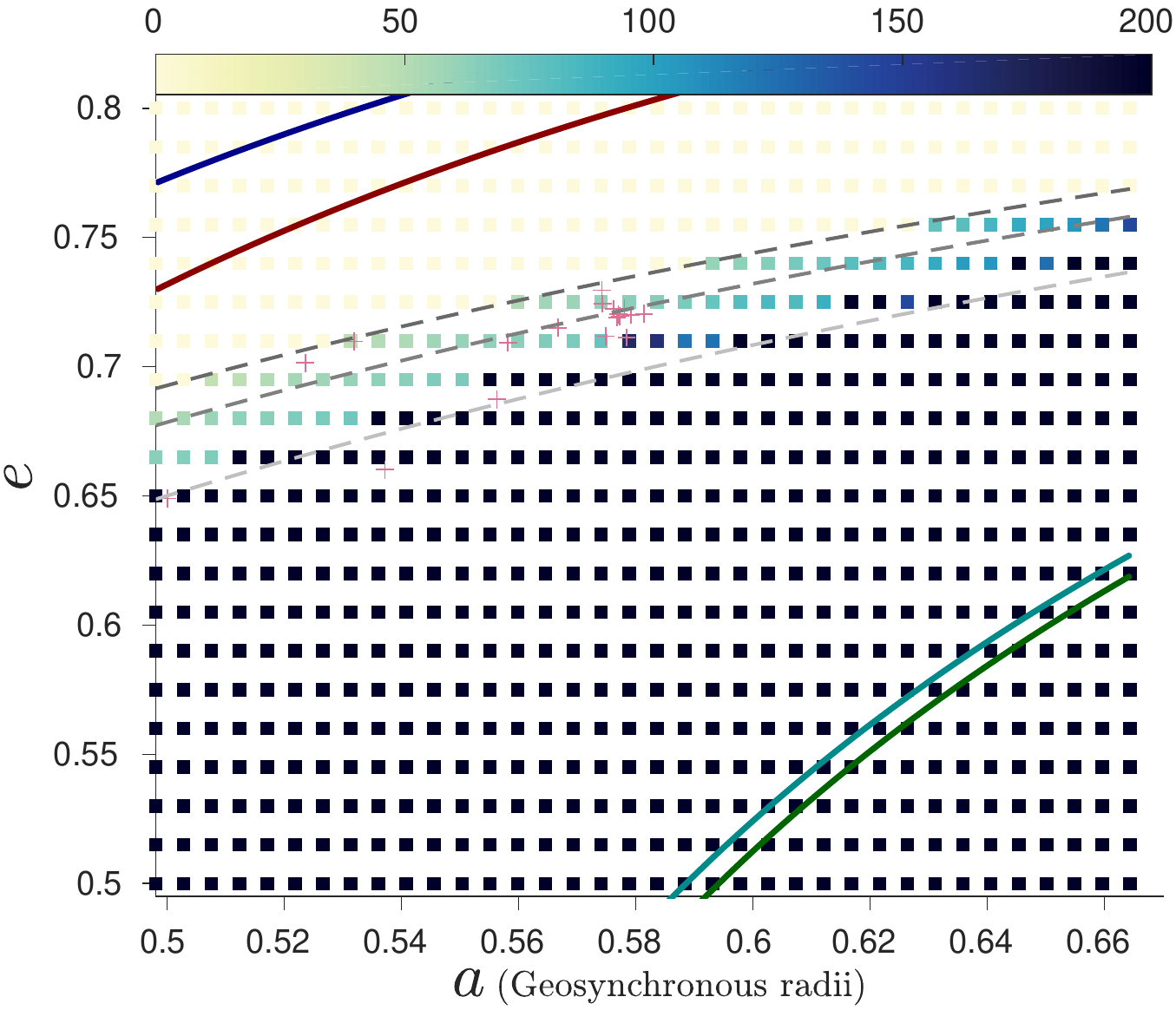} &
  \includegraphics[width=6.0cm,height=4.75cm]{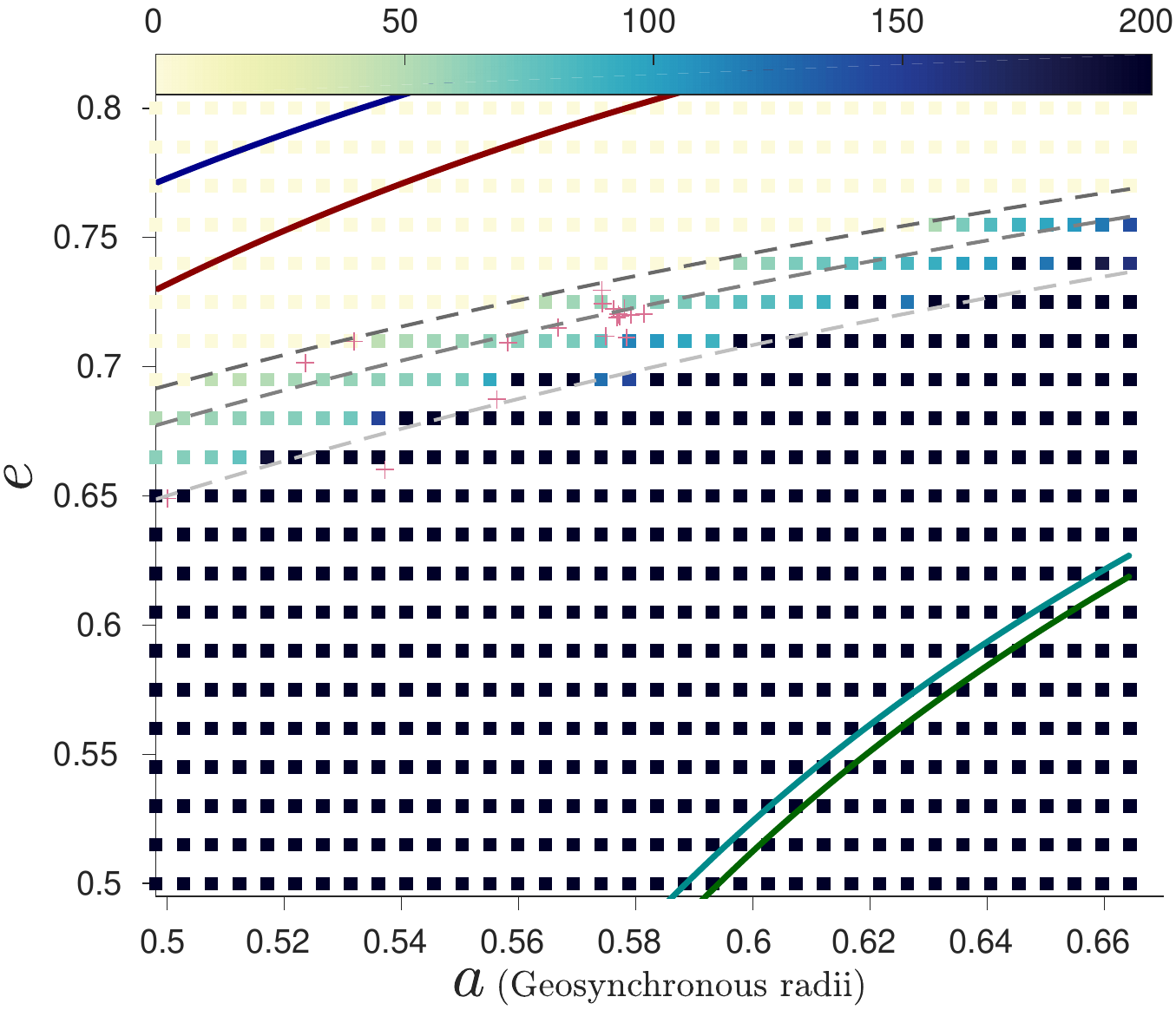}\\
  \multicolumn{2}{c} {$\bf i_{o} = 63.4^\circ$ }\\
  \includegraphics[width=6.0cm,height=4.75cm]{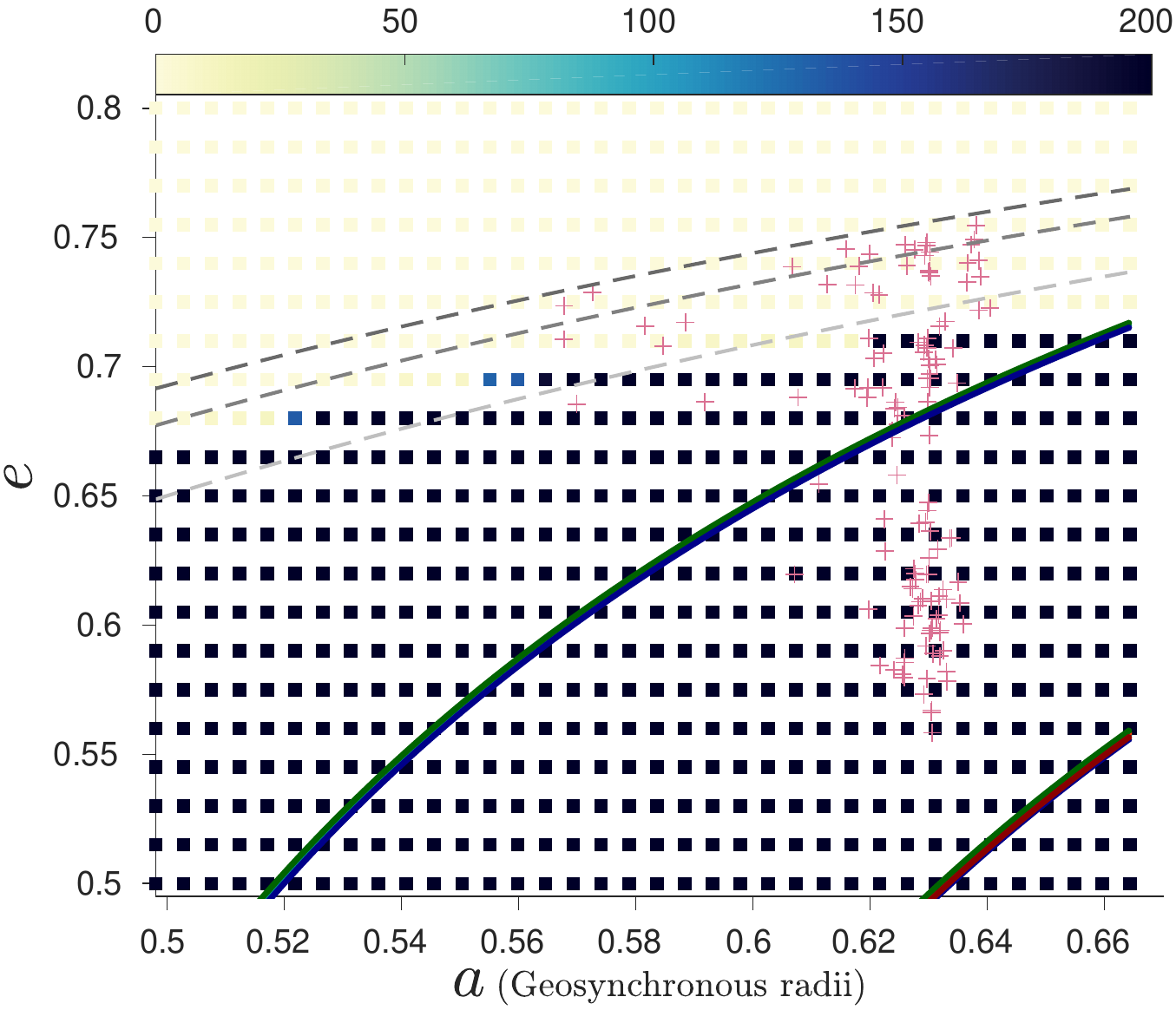} &
  \includegraphics[width=6.0cm,height=4.75cm]{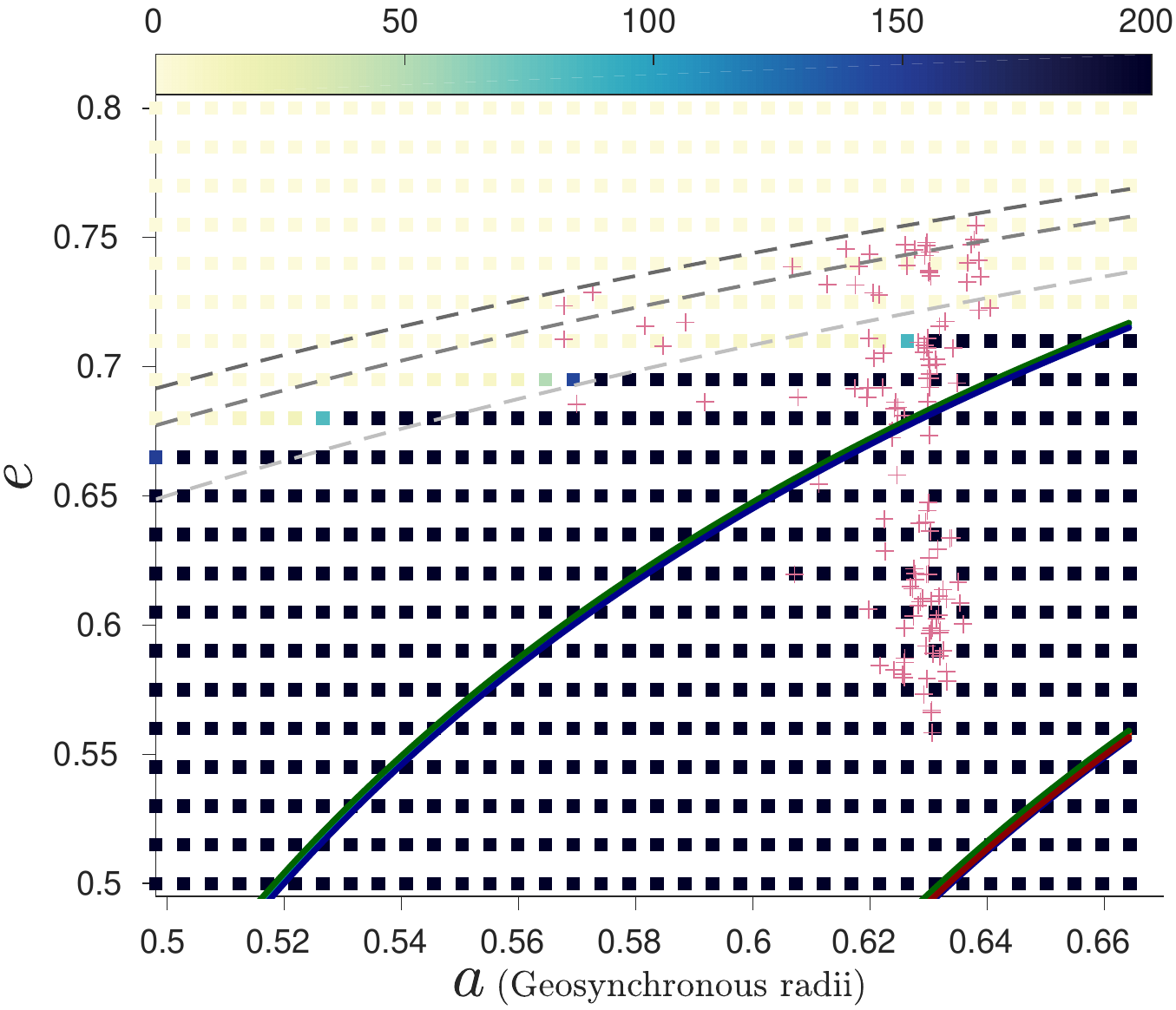}\\
\end{tabular}
\caption{Lifetime maps of the GTO region in $a-e$  for different initial $i$, and $\Delta\Omega = 90^\circ$, $\Delta\omega = 270^\circ$, Epoch 2018, $A/m = 
0.02$ m$^2$/kg, with atmospheric drag (right) or not (left).
  The red points mark the cataloged population with $i=i_{o} \pm 5^{\circ}$ and $A/m\in\left[0:0.02\right]$ m$^2$/kg. 
  The colorbar goes from lifetime 0 to 200 years.  The gray dashed curves correspond to $~100,~400,~1000\ km$ initial altitude ($q_0$, from high to low $e$). The multicolored bold lines correspond to the strongest lunisolar resonances that are present for each inclination. See text for more details.}
\label{fig:lifemaps_conf8_srp1}      
\end{figure}

\begin{figure}[htp!]
	\captionsetup{justification=justified}
	\centering 
\begin{tabular}{cc}
  \textbf{NO DRAG} & \textbf{DRAG}\\
  \multicolumn{2}{c} {$\bf i_{o} = 5.235^\circ$}\\
  \includegraphics[width=6.0cm,height=4.75cm]{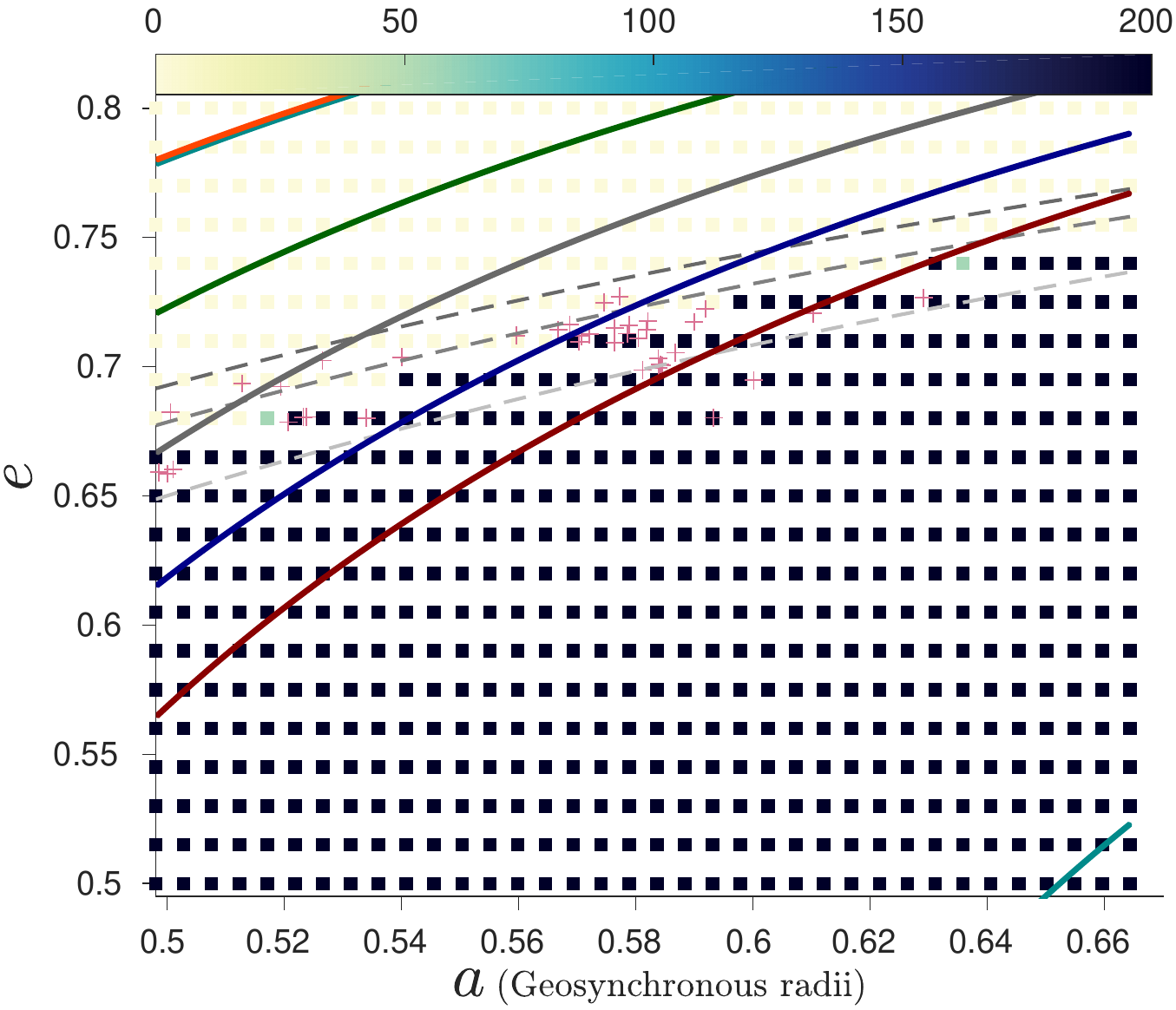} &
  \includegraphics[width=6.0cm,height=4.75cm]{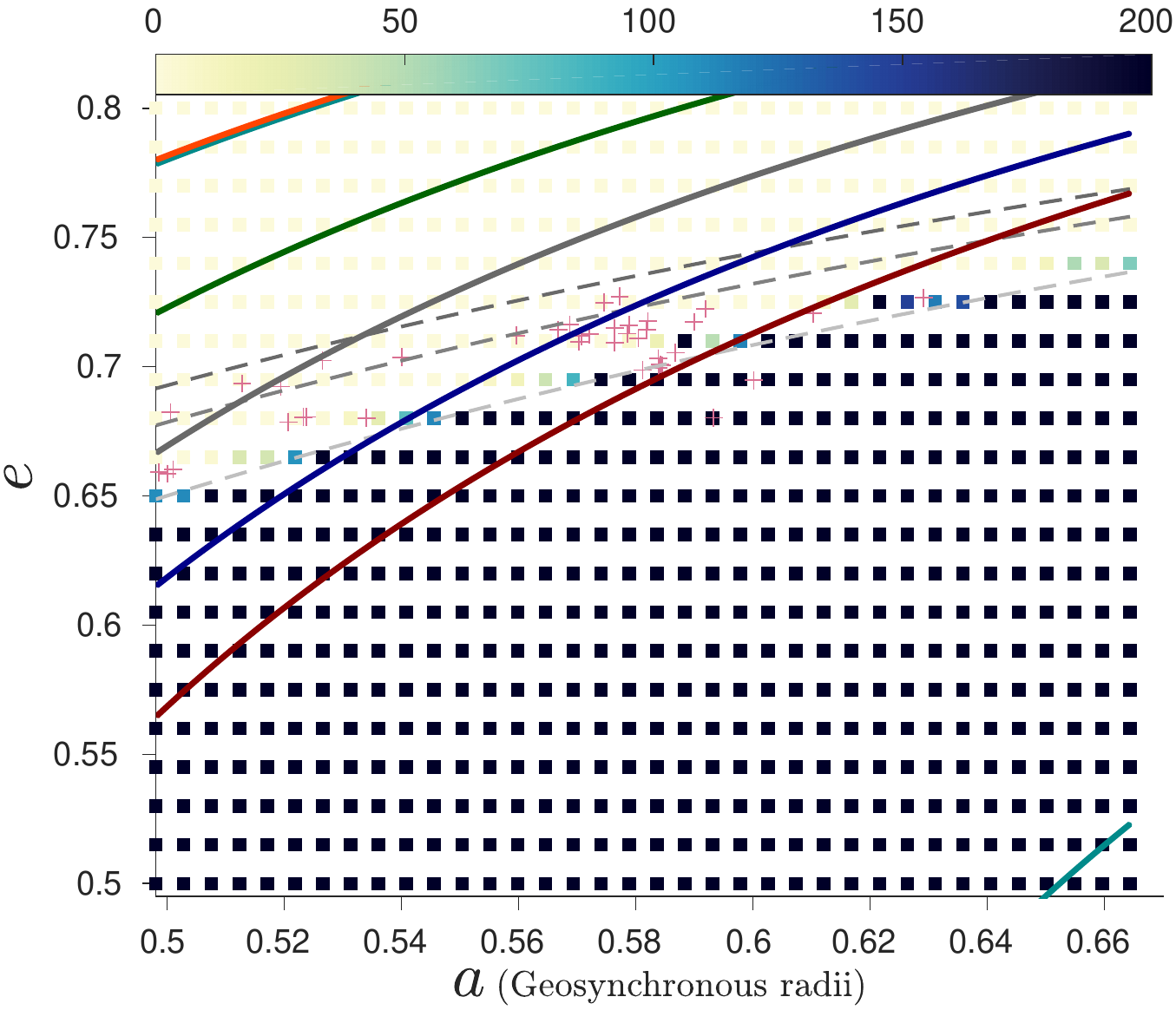} \\
  \multicolumn{2}{c} {$\bf i_{o} = 28.533^\circ$}\\
  \includegraphics[width=6.0cm,height=4.75cm]{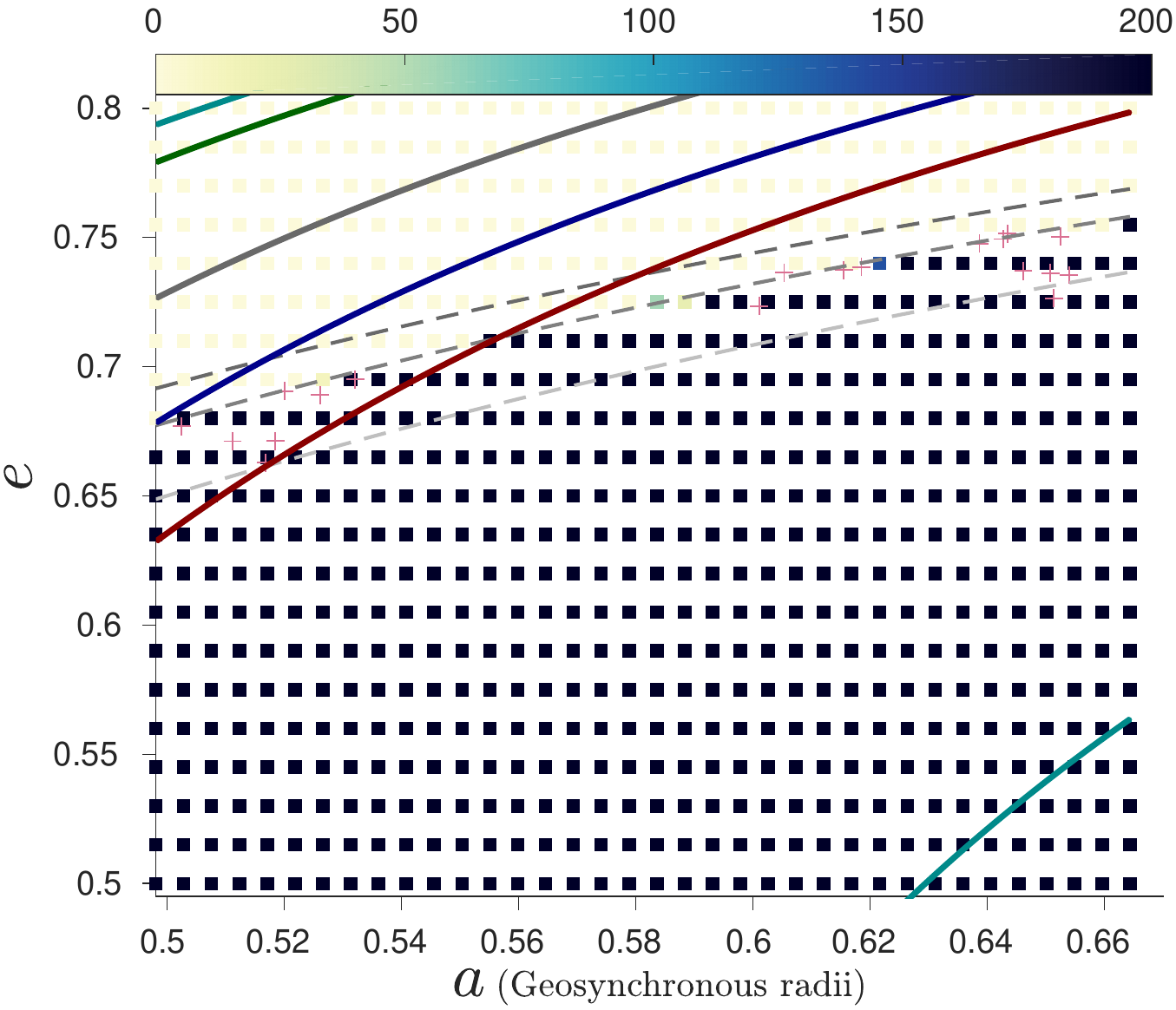} &
  \includegraphics[width=6.0cm,height=4.75cm]{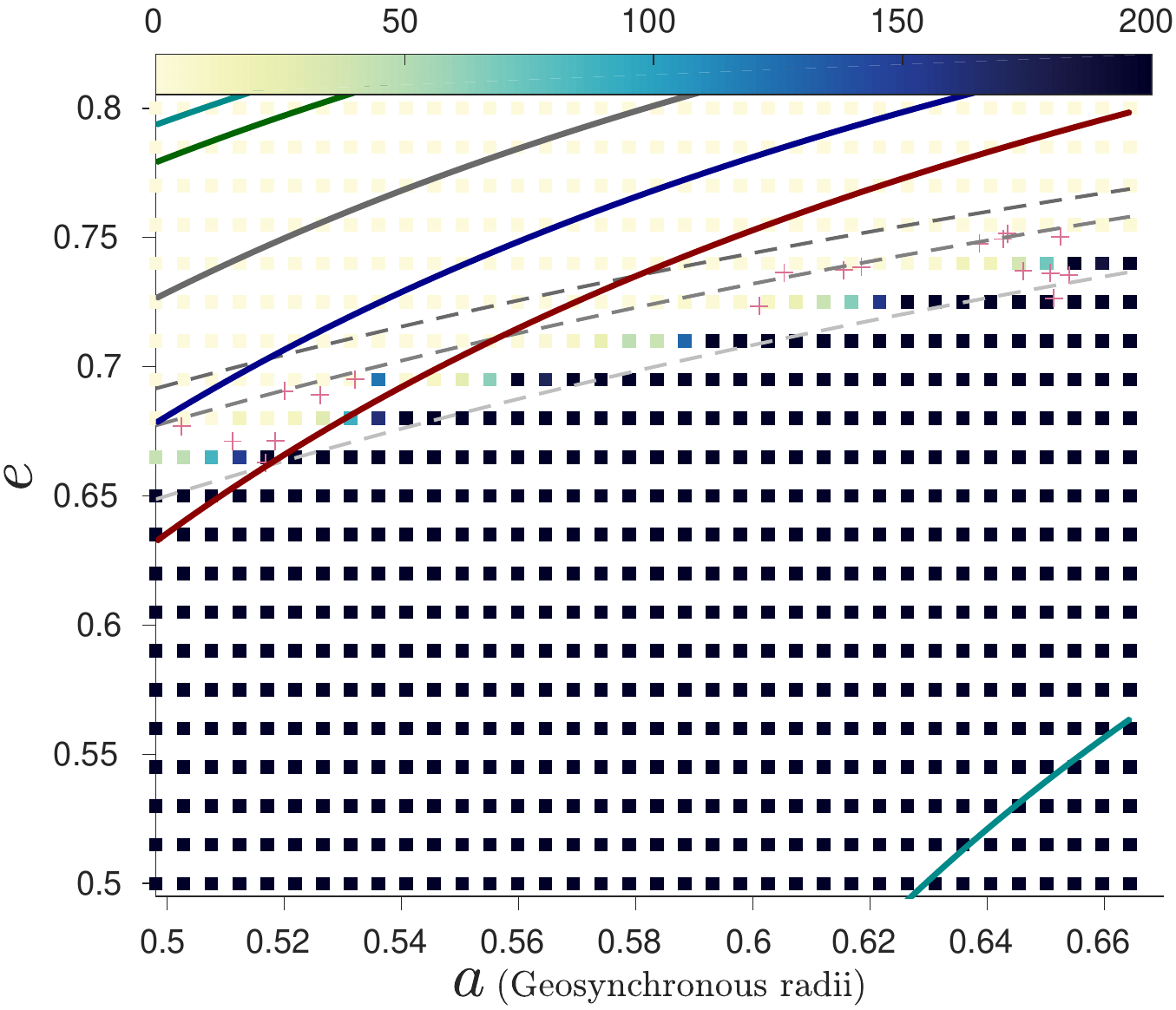}\\
  \multicolumn{2}{c} {$\bf i_{o} = 46^\circ$}\\
  \includegraphics[width=6.0cm,height=4.75cm]{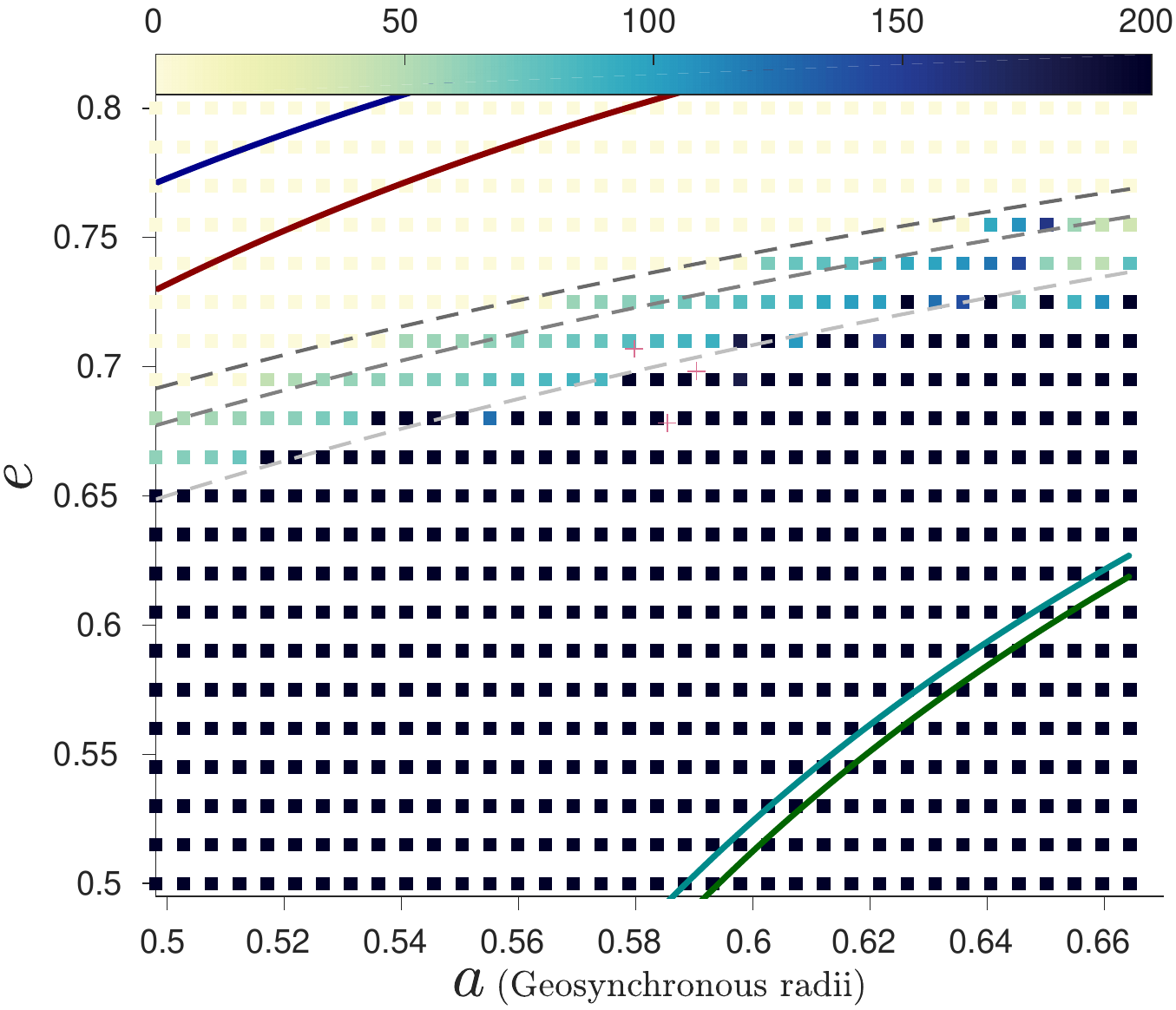} &
  \includegraphics[width=6.0cm,height=4.75cm]{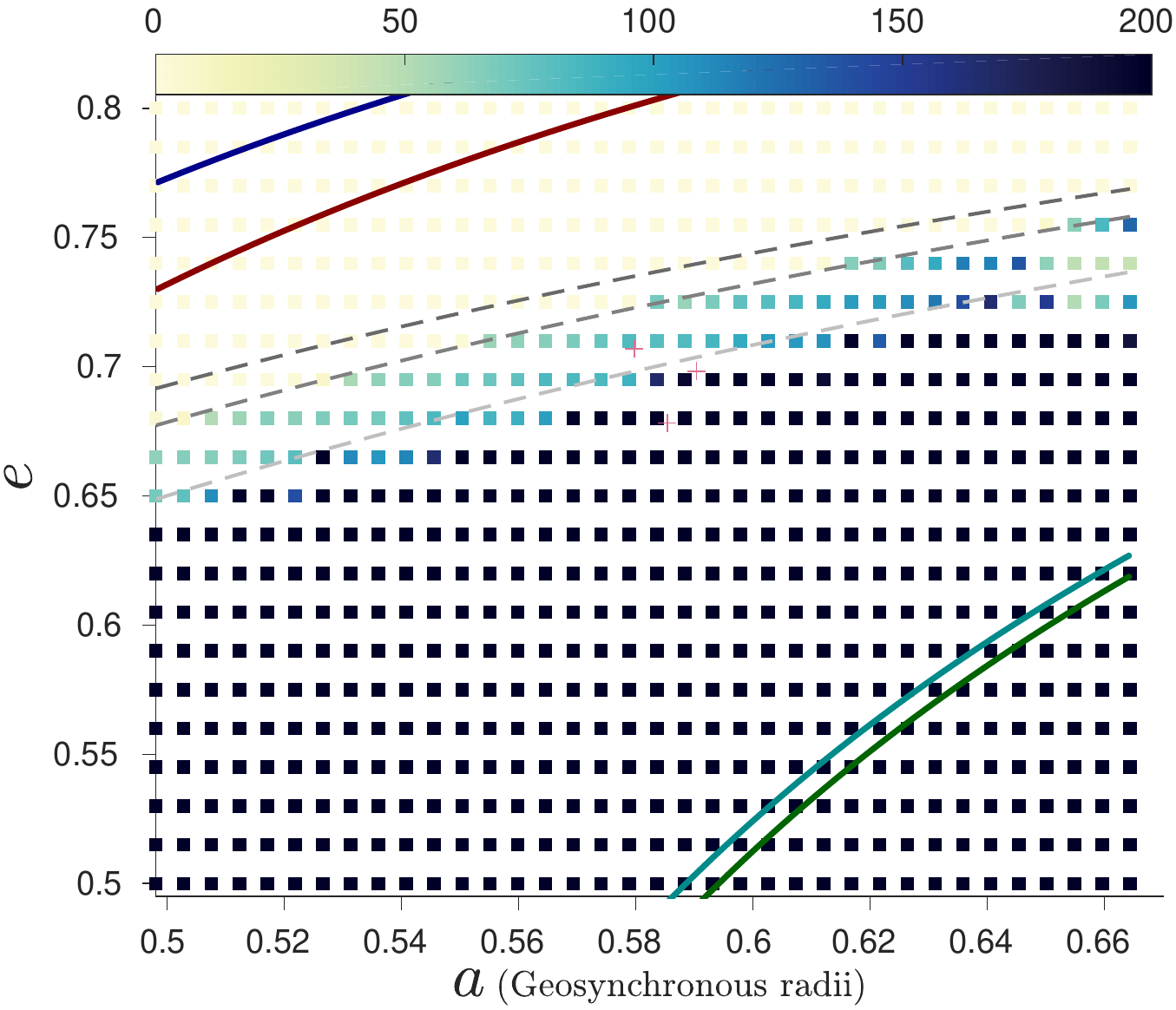}\\
  \multicolumn{2}{c} {$\bf i_{o} = 63.4^\circ$ }\\
  \includegraphics[width=6.0cm,height=4.75cm]{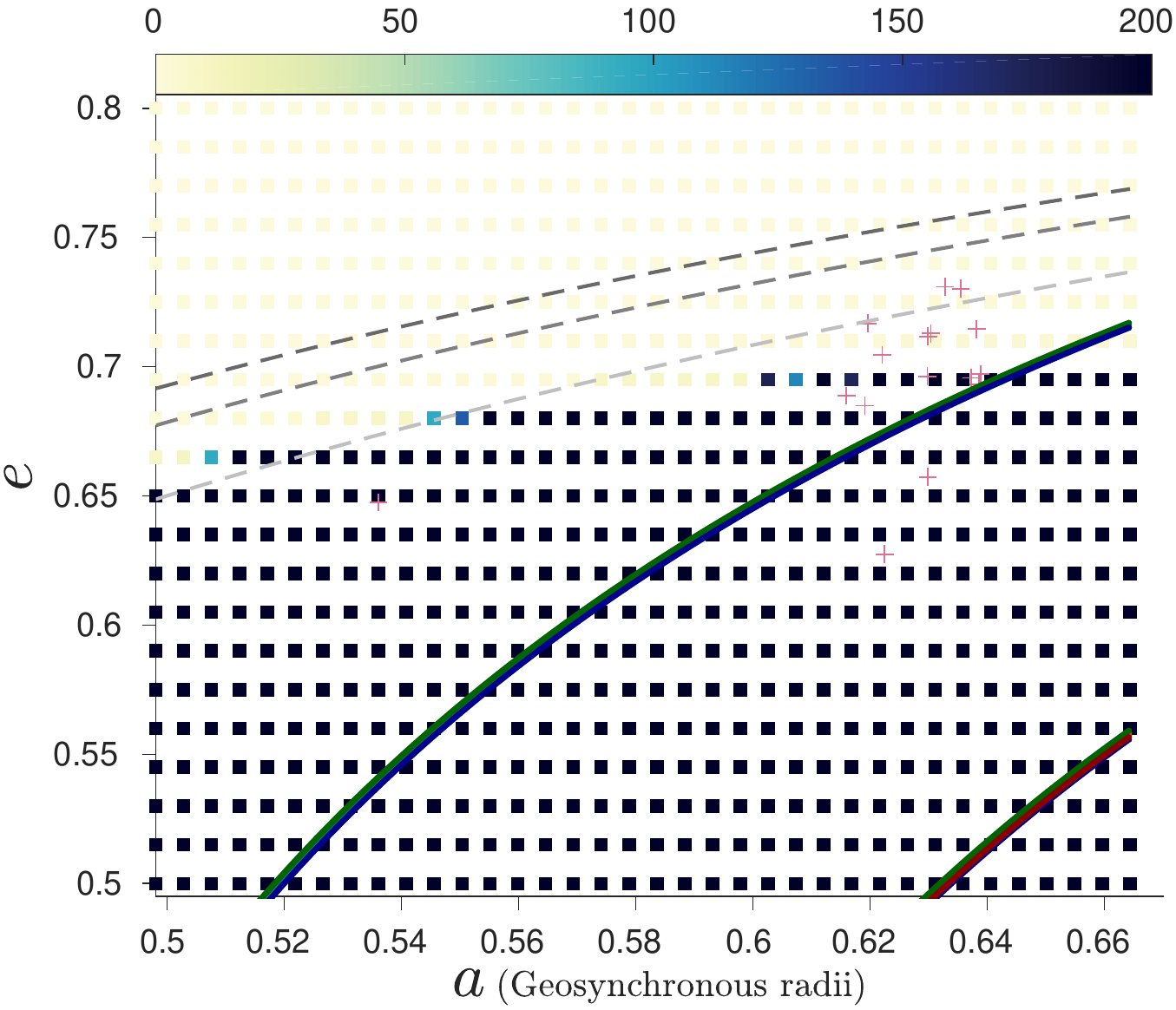} &
  \includegraphics[width=6.0cm,height=4.75cm]{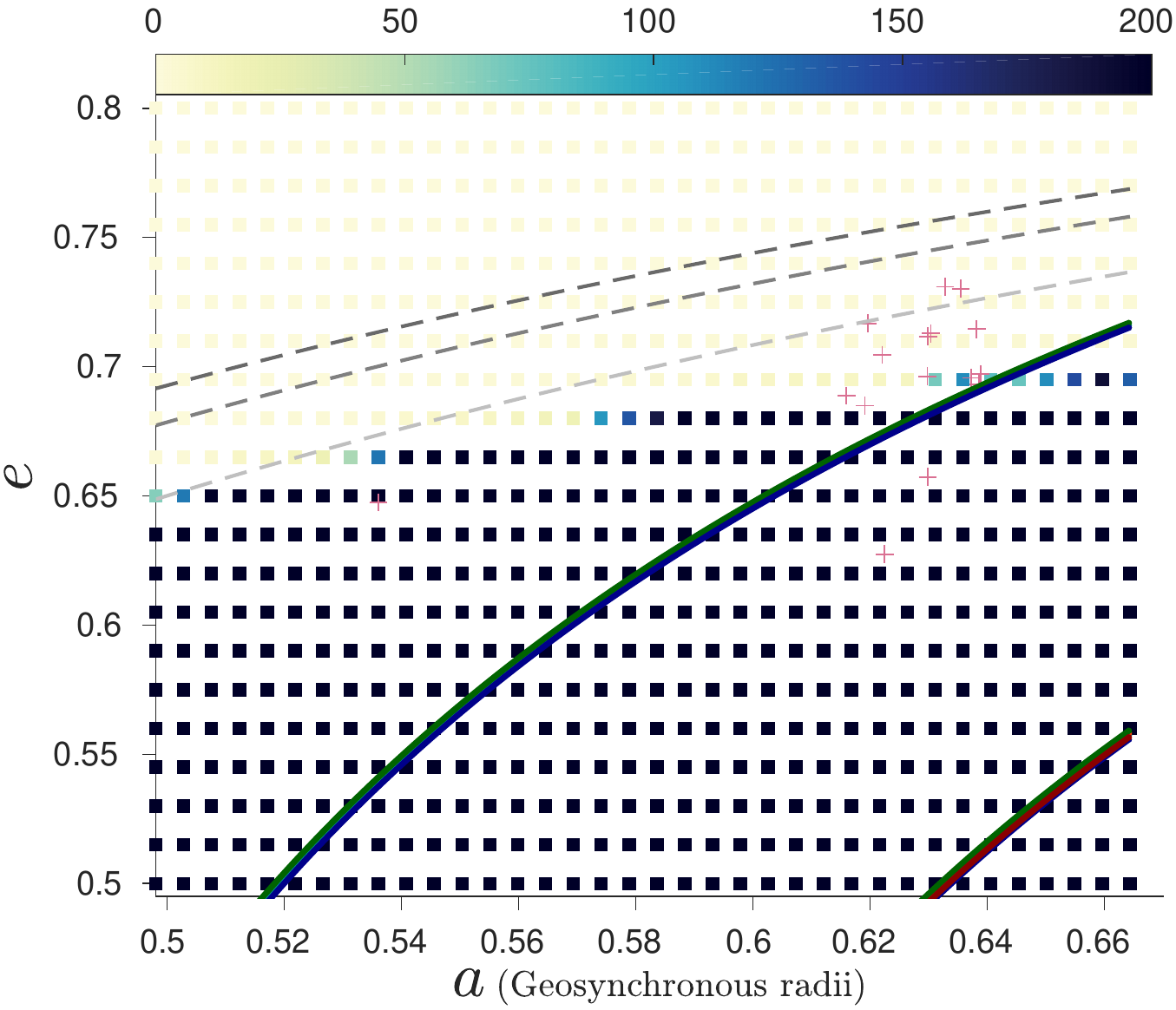}\\
\end{tabular}
\caption{Lifetime maps of the GTO region in $a-e$  for different initial $i$, and $\Delta\Omega = 90^\circ$, $\Delta\omega = 270^\circ$, Epoch 2018, $A/m = 1$ m$^2$/kg, with atmospheric drag (right) or not (left). The red points mark the cataloged population with $i=i_{o} \pm 5^{\circ}$ and $A/m\in\left(\left.0.02:1\right.\right]$ m$^2$/kg. 
  The colorbar goes from lifetime 0 to 200 years.  The gray dashed curves correspond to $~100,~400,~1000\ km$ initial altitude ($q_0$, from high to low $e$). The multicolored bold lines correspond to the strongest lunisolar resonances that are present for each inclination. See text for more details.}
\label{fig:lifemaps_conf8_srp2}      
\end{figure}

\begin{figure}[htp!]
	\captionsetup{justification=justified}
	\centering 
\begin{tabular}{cc}
  \textbf{NO DRAG} & \textbf{DRAG}\\
  \multicolumn{2}{c} {$\bf i_{o} = 5.235^\circ$}\\
  \includegraphics[width=6.0cm,height=4.75cm]{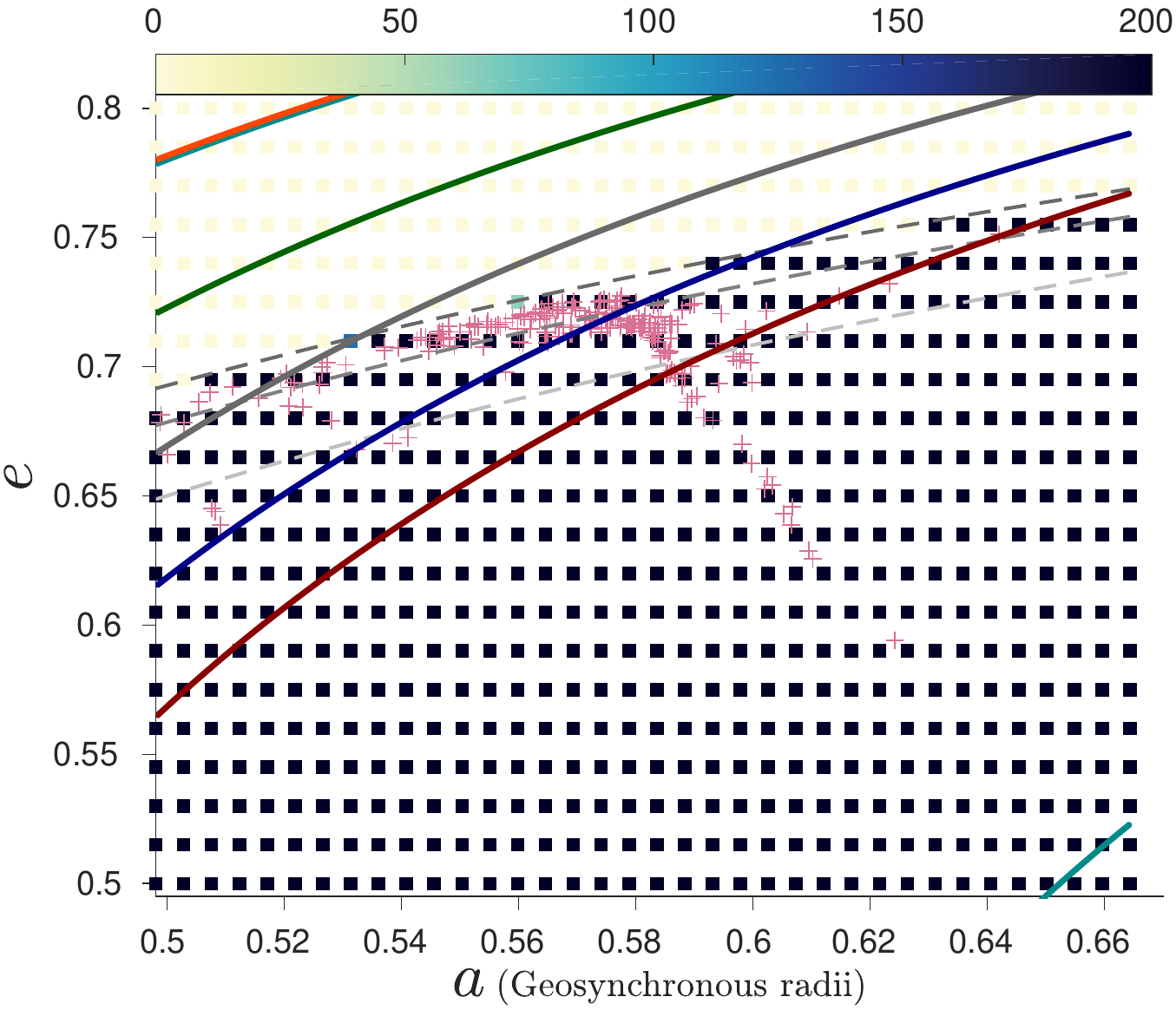} &
  \includegraphics[width=6.0cm,height=4.75cm]{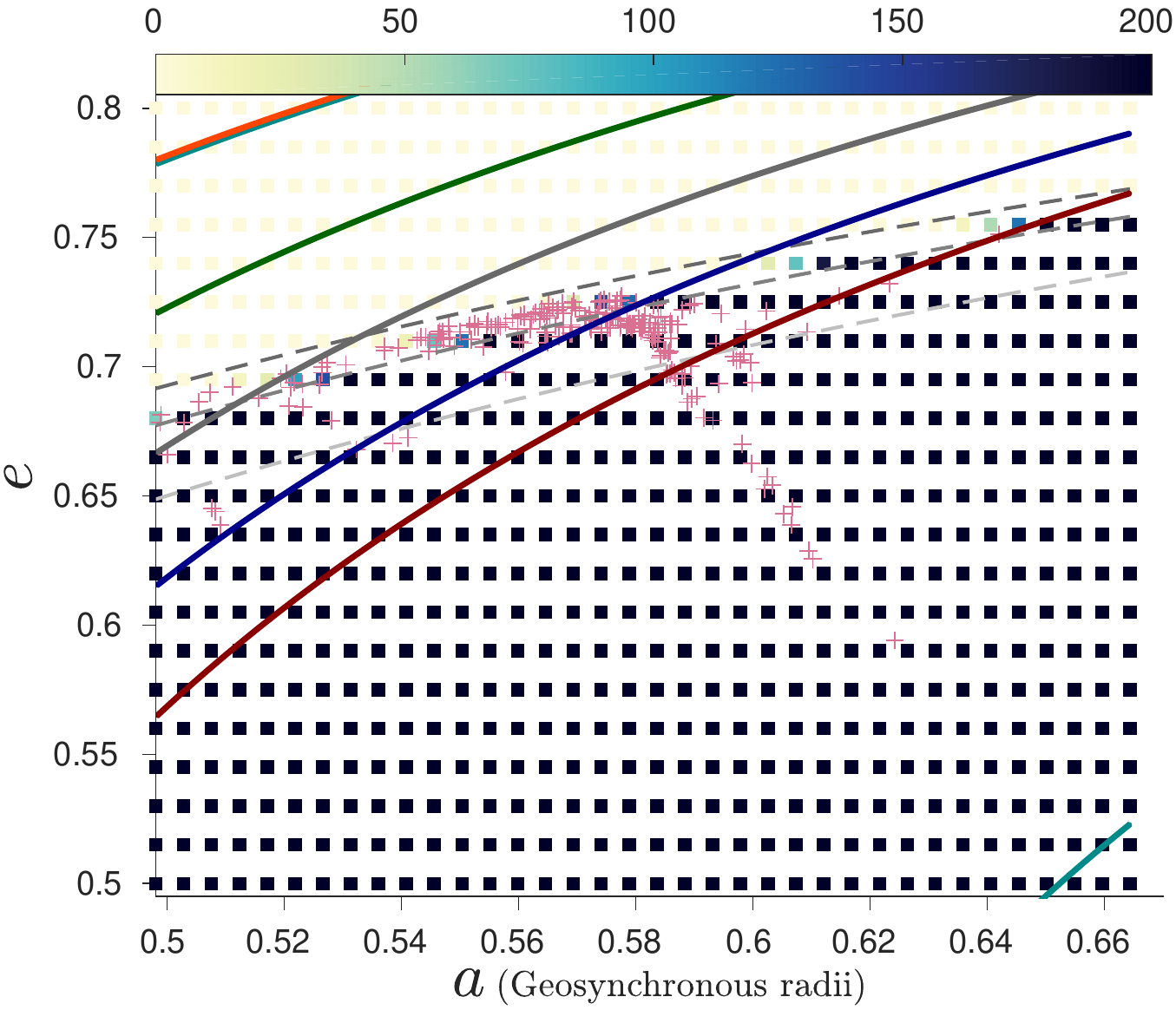} \\
  \multicolumn{2}{c} {$\bf i_{o} = 28.533^\circ$}\\
  \includegraphics[width=6.0cm,height=4.75cm]{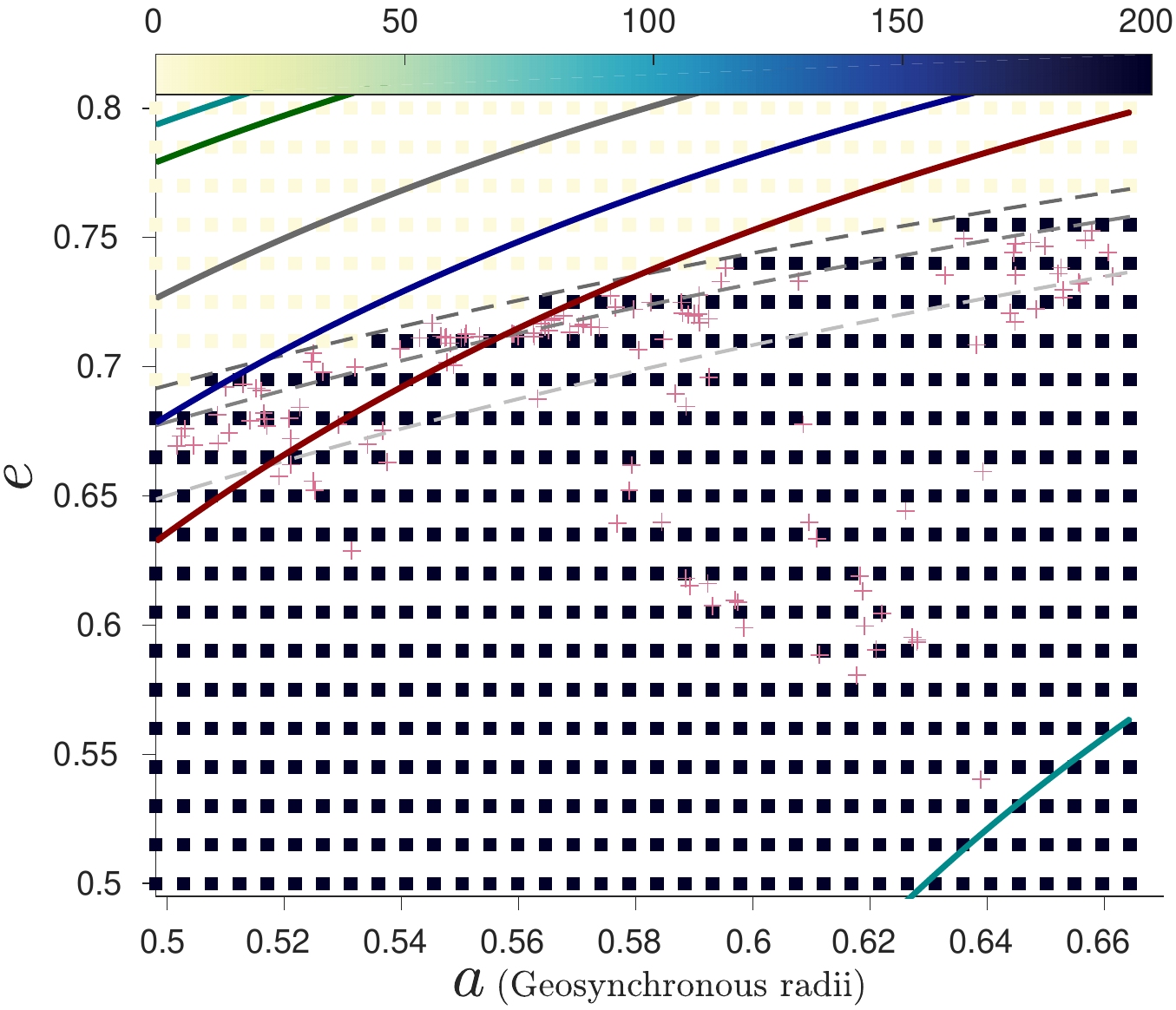} &
  \includegraphics[width=6.0cm,height=4.75cm]{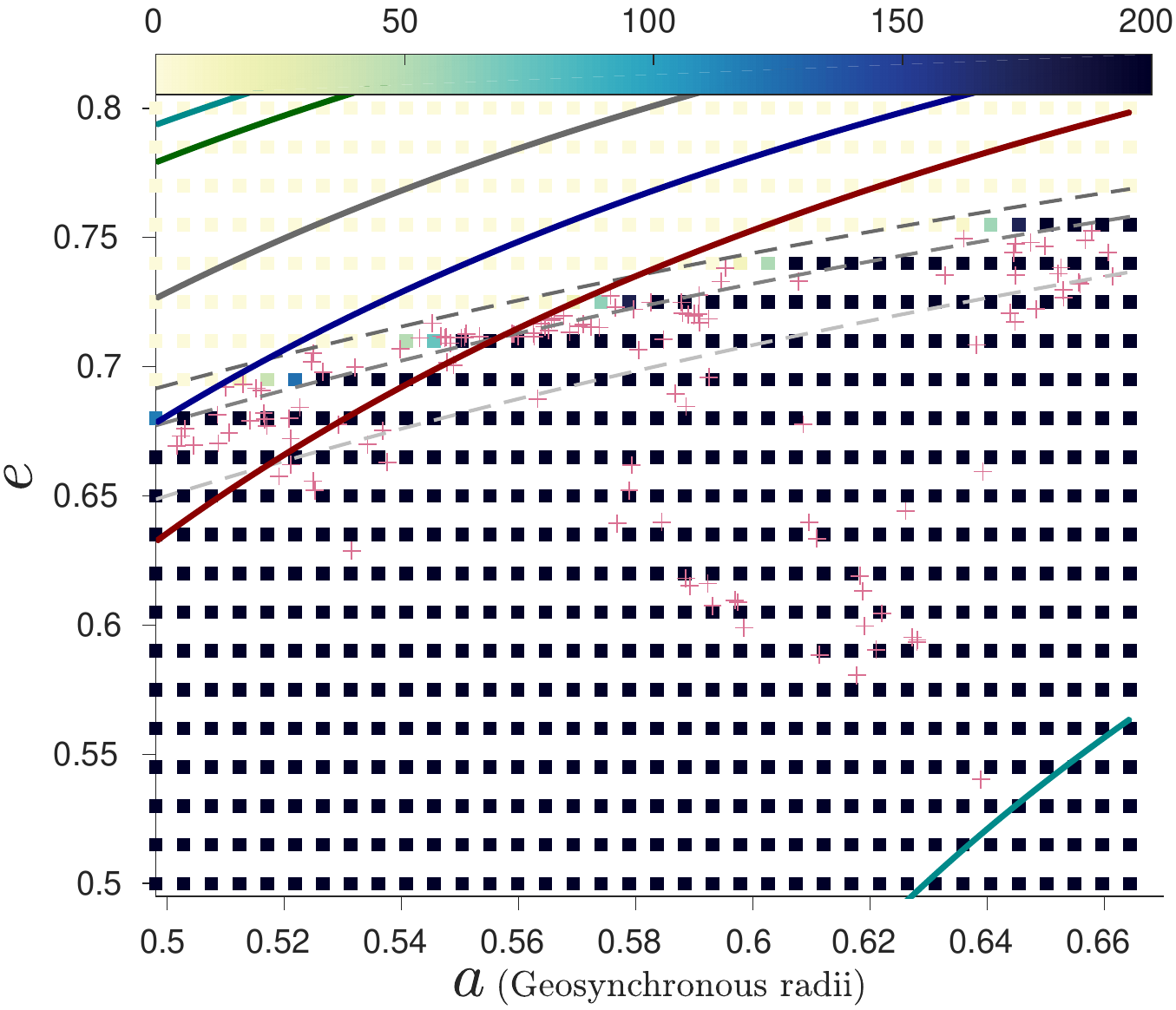}\\
  \multicolumn{2}{c} {$\bf i_{o} = 46^\circ$}\\
  \includegraphics[width=6.0cm,height=4.75cm]{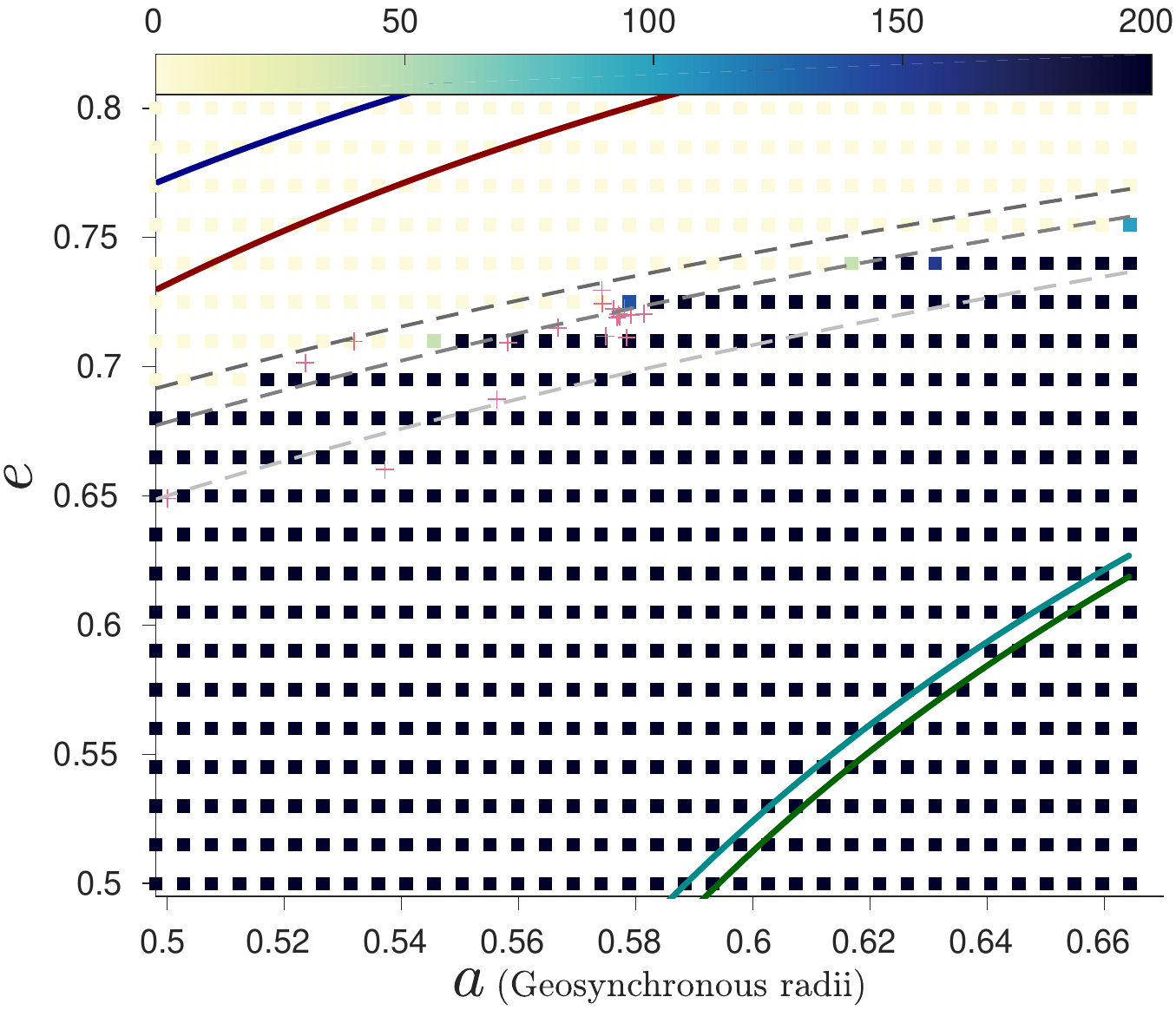} &
  \includegraphics[width=6.0cm,height=4.75cm]{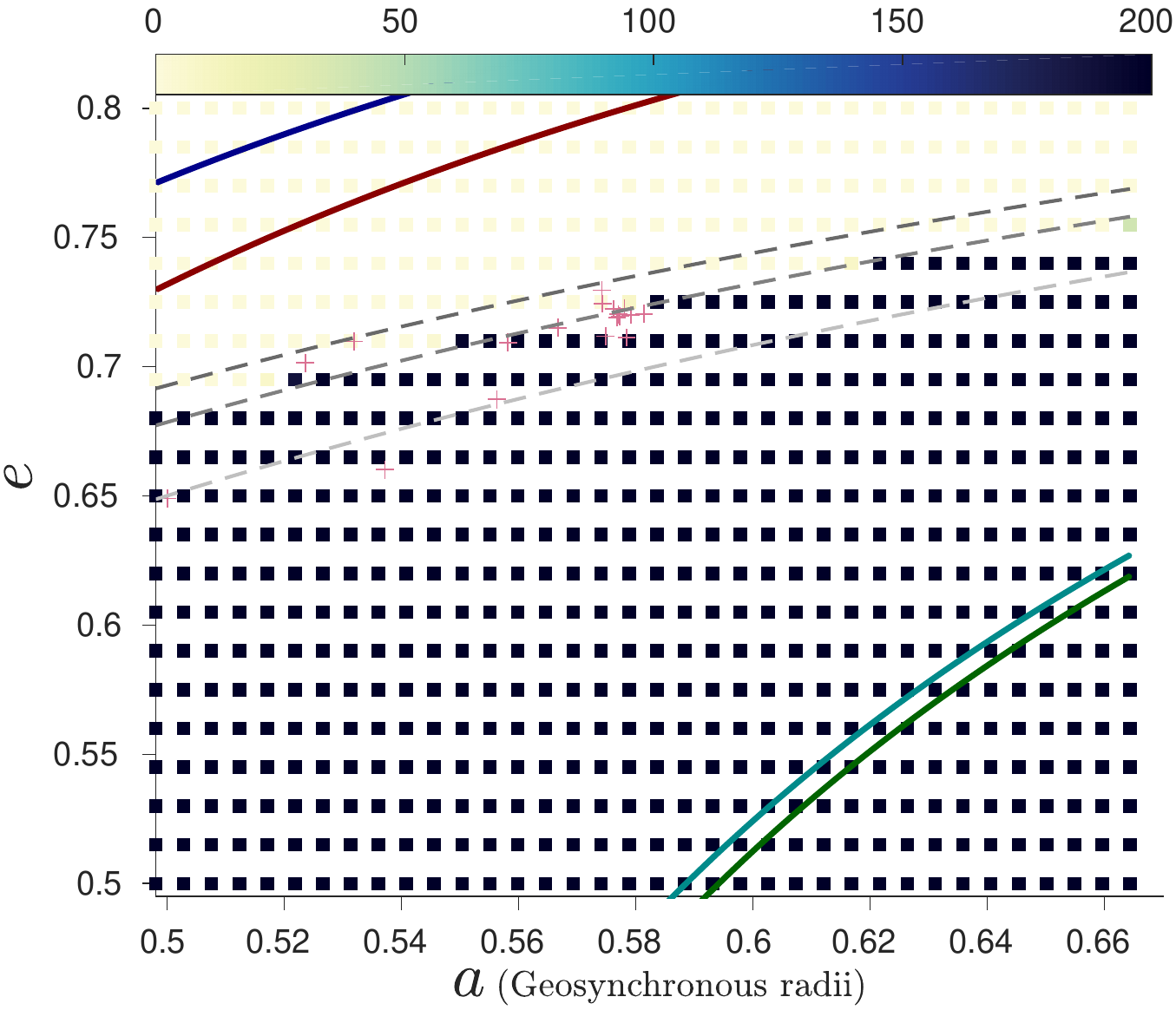}\\
  \multicolumn{2}{c} {$\bf i_{o} = 63.4^\circ$ }\\
  \includegraphics[width=6.0cm,height=4.75cm]{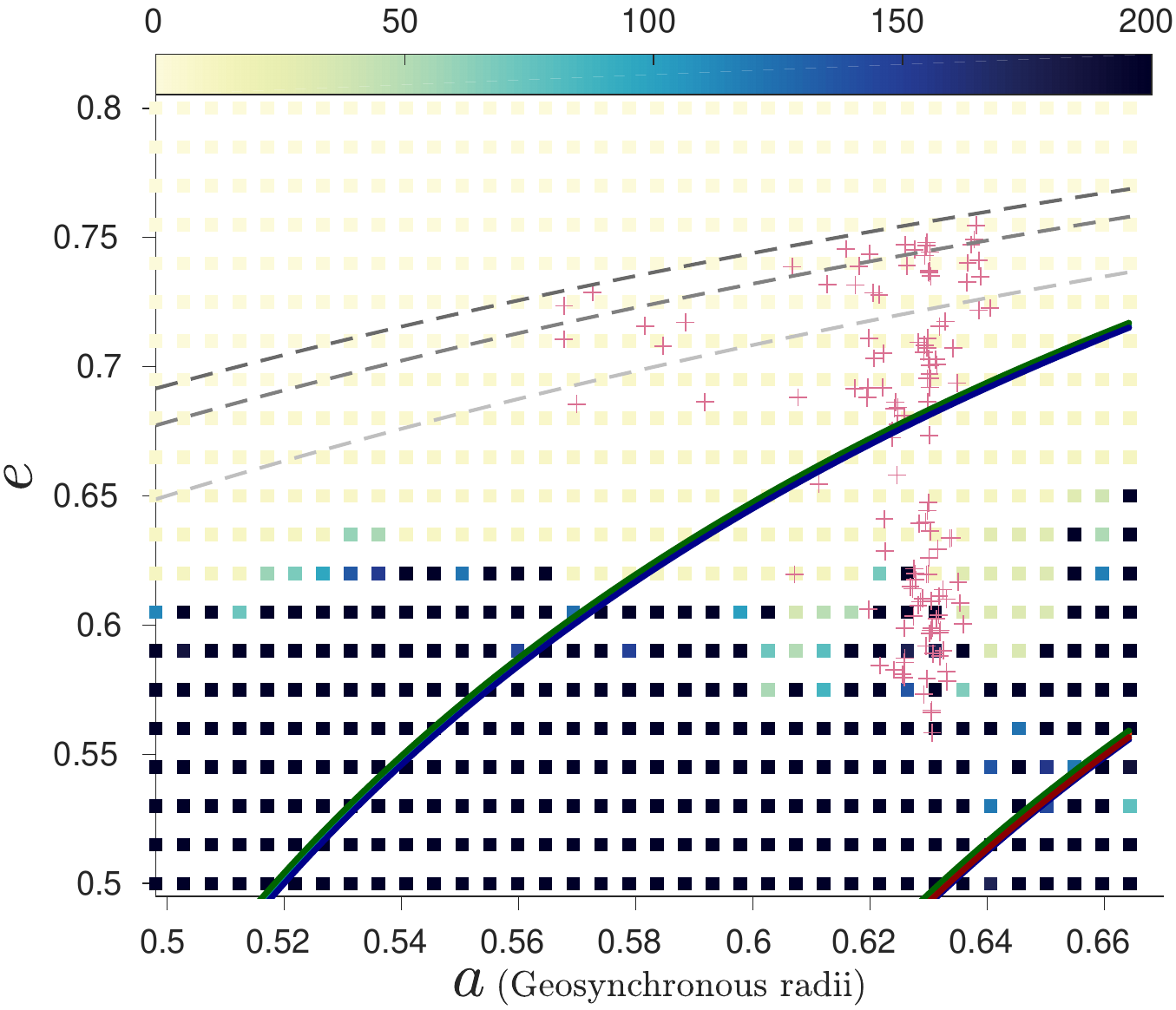} &
  \includegraphics[width=6.0cm,height=4.75cm]{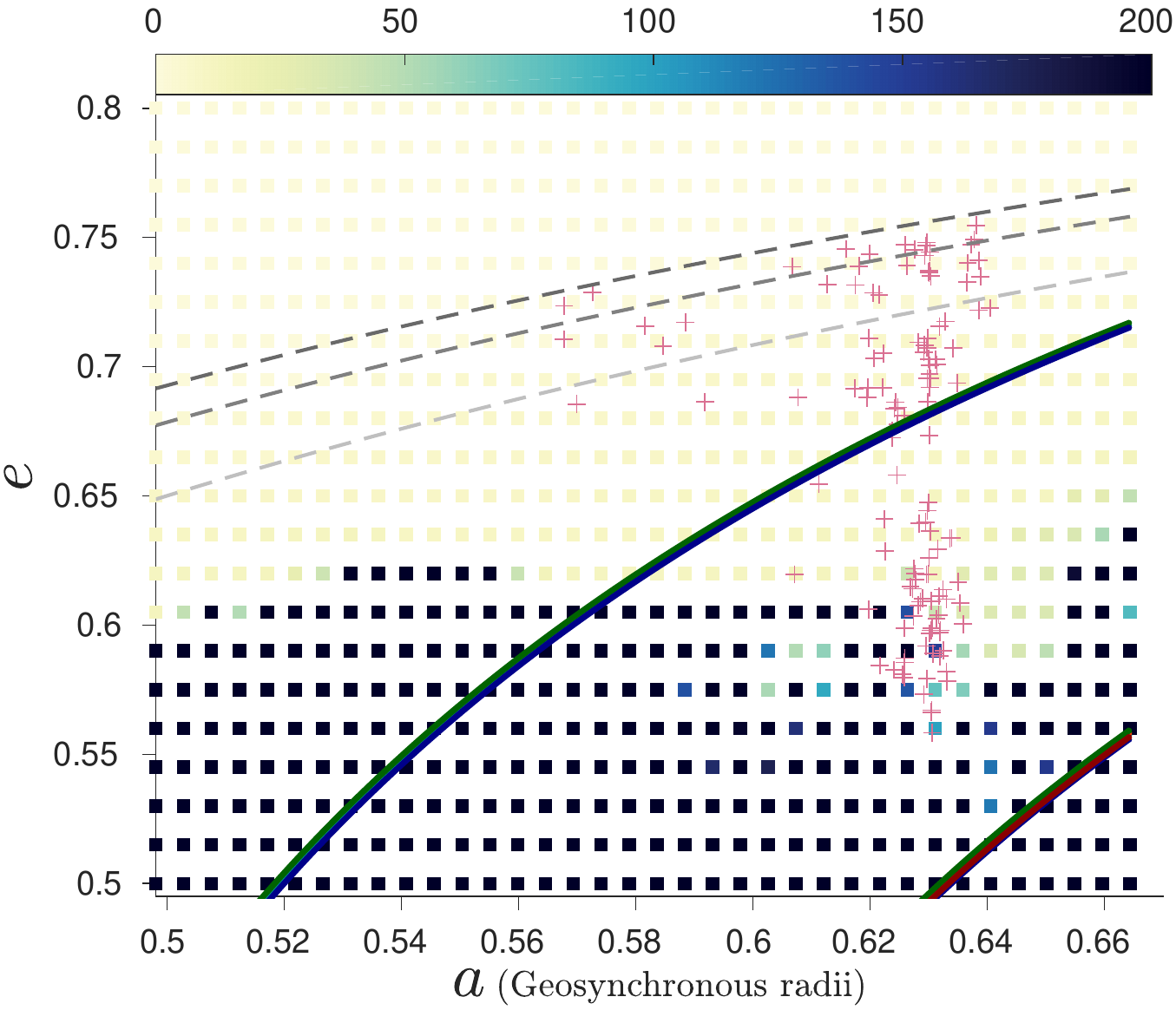}\\
\end{tabular}
\caption{Lifetime maps of the GTO region in $a-e$  for different initial $i$, and $\Delta\Omega = 180^\circ$, $\Delta\omega = 270^\circ$, Epoch 2020, $A/m = 
0.02$ m$^2$/kg, with atmospheric drag (right) or not (left).  The red points mark the cataloged population with $i=i_{o} \pm 5^{\circ}$ and $A/m\in\left[0:0.02\right]$ m$^2$/kg. The colorbar goes from lifetime 0 to 200 years.  The gray dashed curves correspond to $~100,~400,~1000\ km$ initial altitude ($q_0$, from high to low $e$). The multicolored bold lines correspond to the strongest lunisolar resonances that are present for each inclination. See text for more details.}
\label{fig:lifemaps_conf12_srp1}      
\end{figure}

\begin{figure}[htp!]
	\captionsetup{justification=justified}
	\centering 
\begin{tabular}{cc}
  \textbf{NO DRAG} & \textbf{DRAG}\\
  \multicolumn{2}{c} {$\bf i_{o} = 5.235^\circ$}\\
  \includegraphics[width=6.0cm,height=4.75cm]{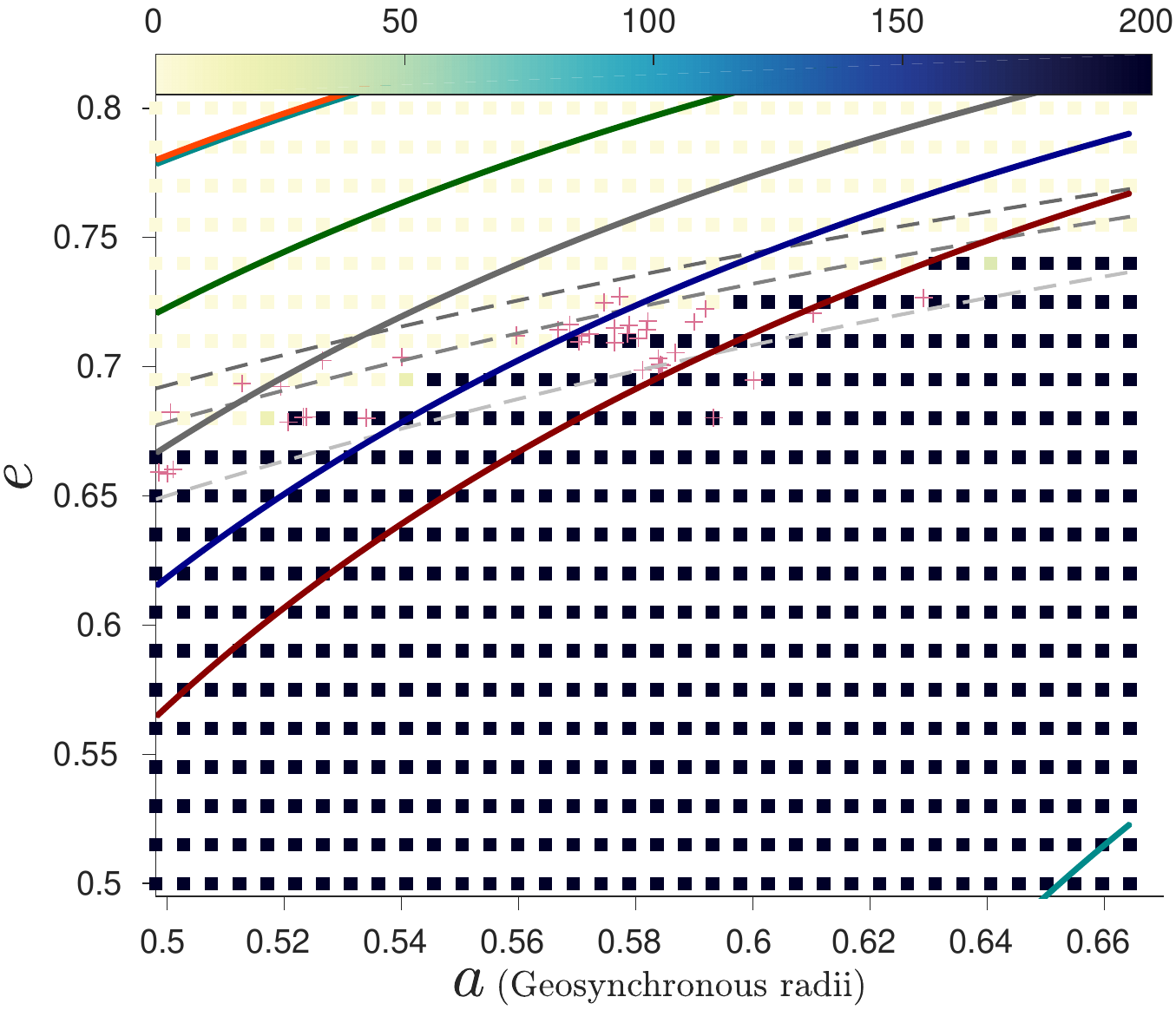} &
  \includegraphics[width=6.0cm,height=4.75cm]{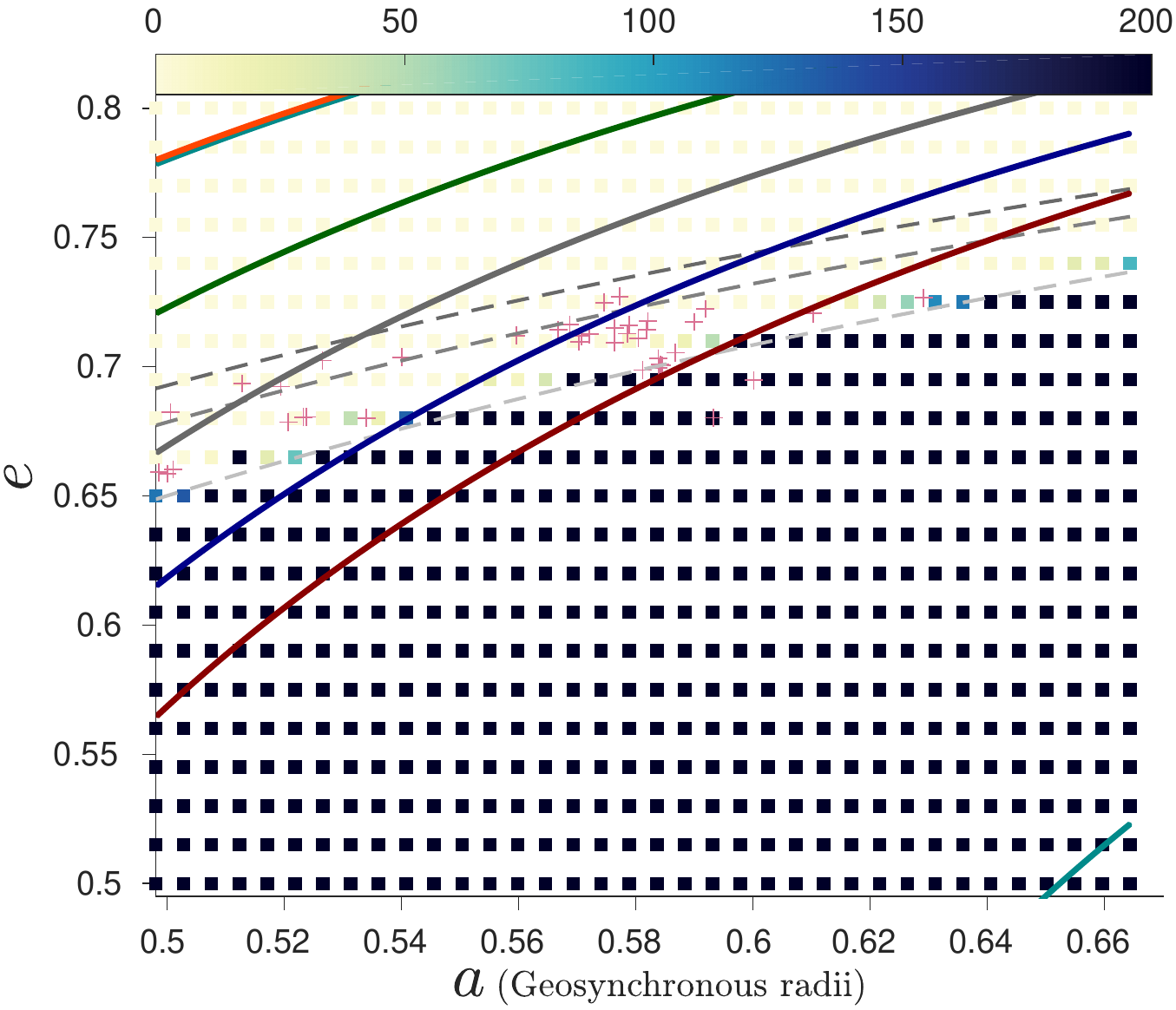} \\
  \multicolumn{2}{c} {$\bf i_{o} = 28.533^\circ$}\\
  \includegraphics[width=6.0cm,height=4.75cm]{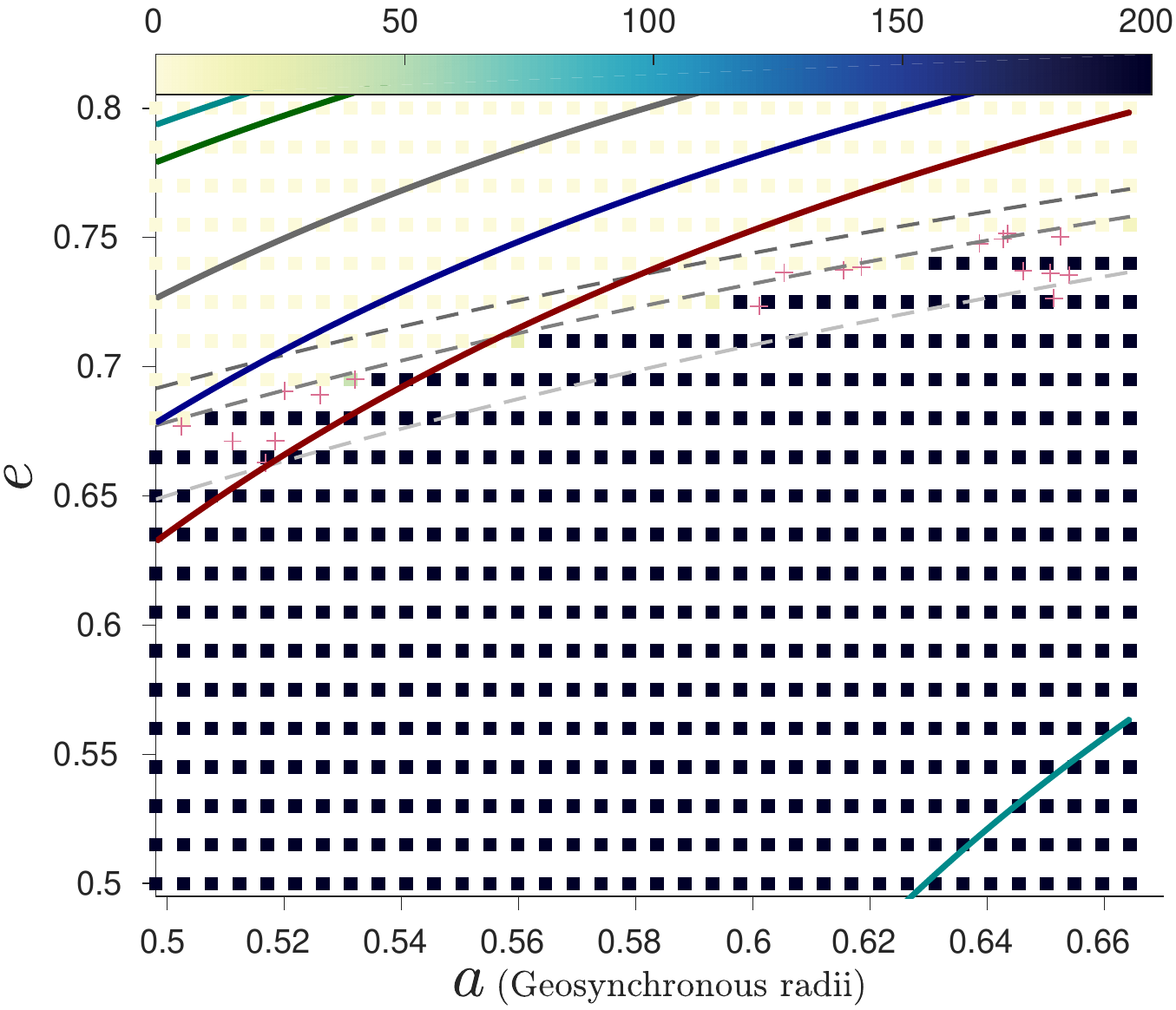} &
  \includegraphics[width=6.0cm,height=4.75cm]{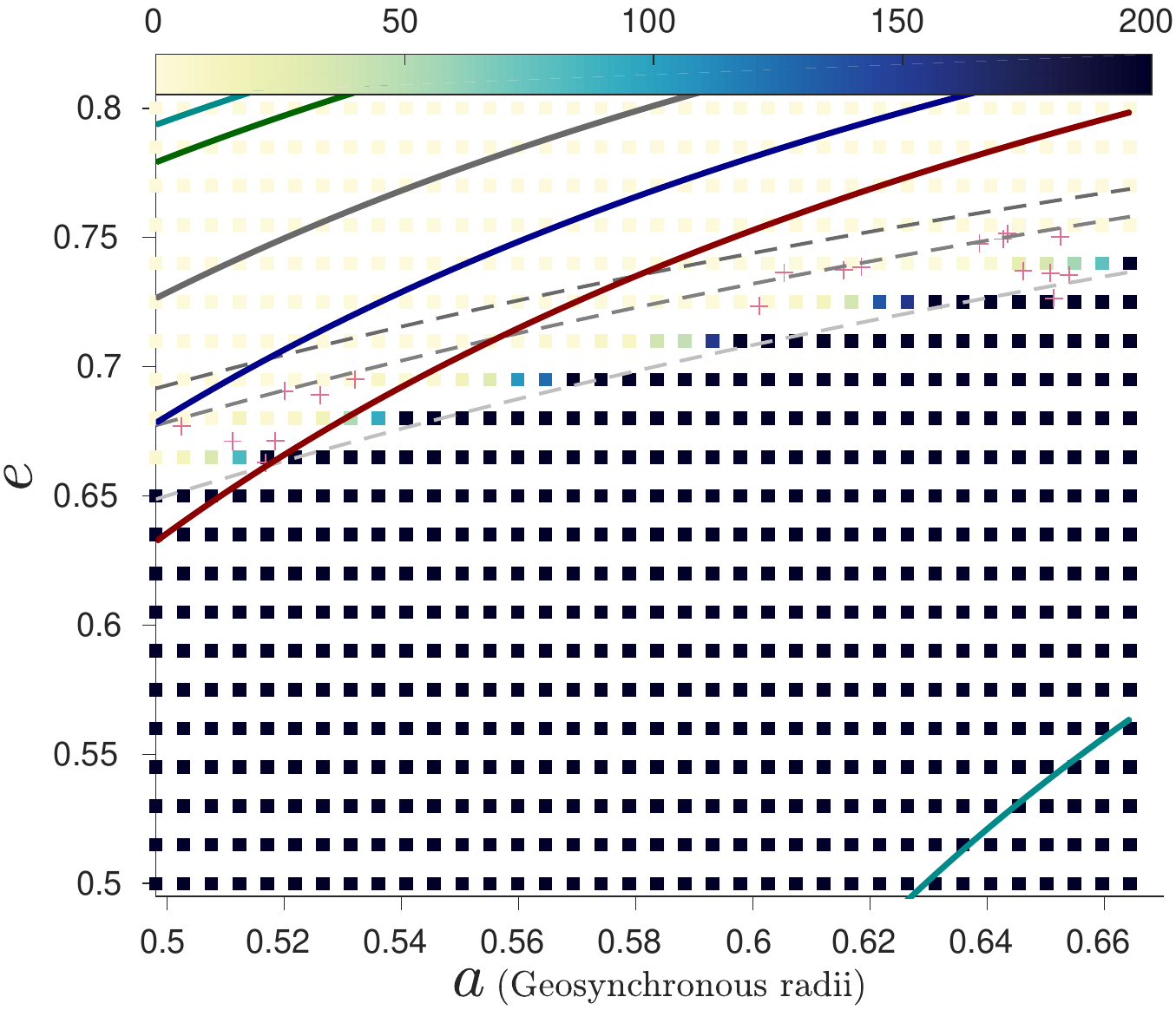}\\
  \multicolumn{2}{c} {$\bf i_{o} = 46^\circ$}\\
  \includegraphics[width=6.0cm,height=4.75cm]{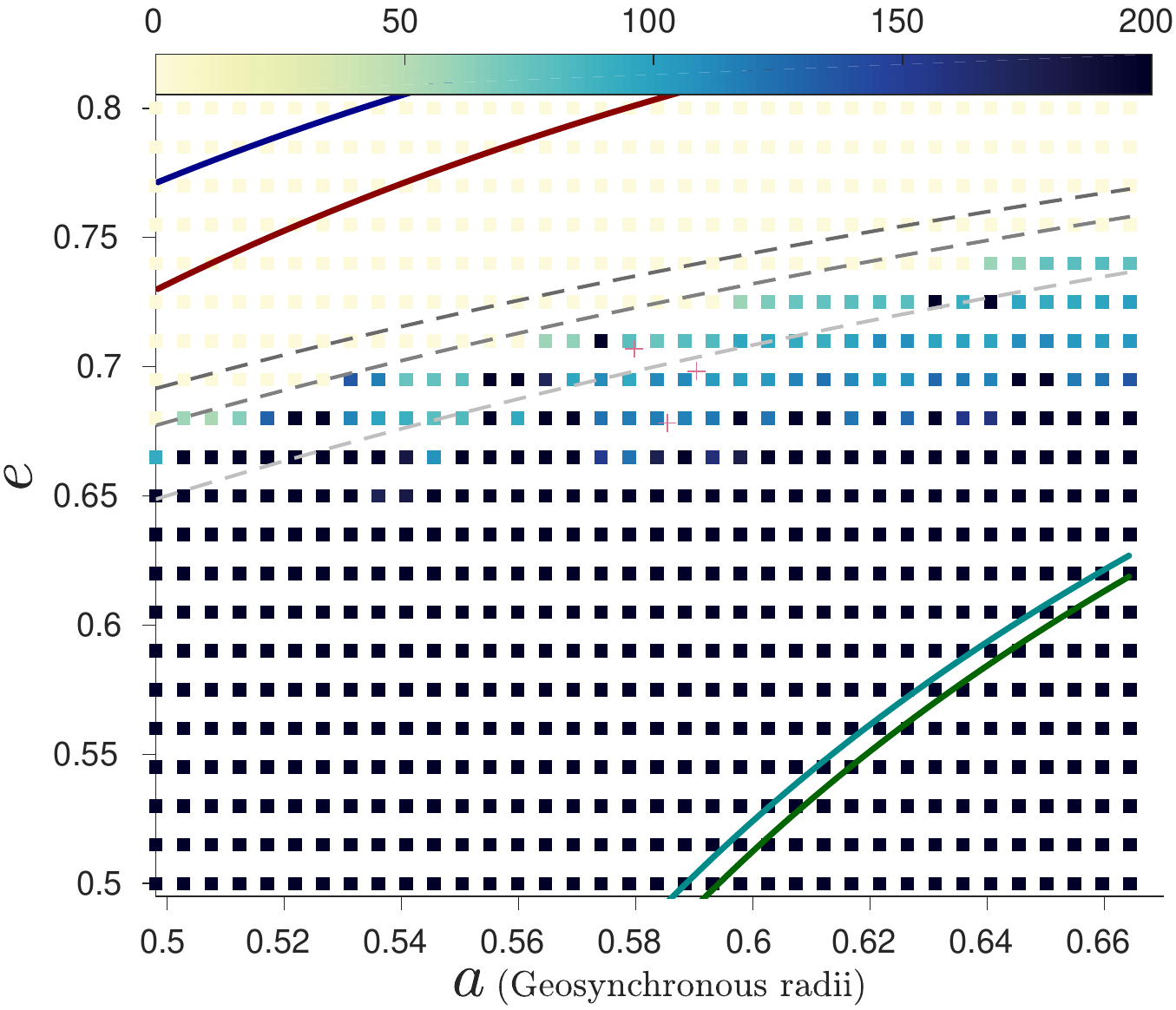} &
  \includegraphics[width=6.0cm,height=4.75cm]{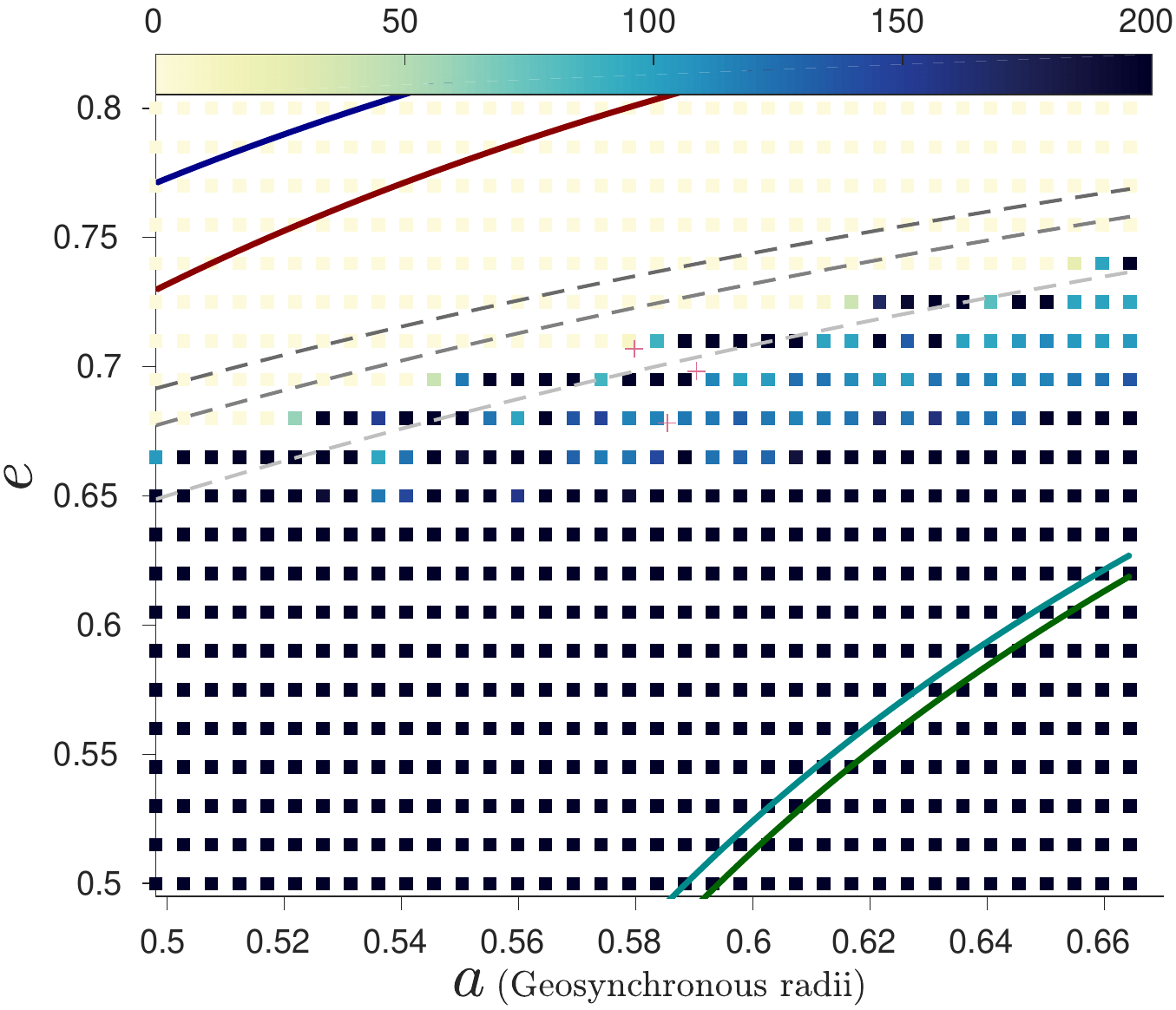}\\
  \multicolumn{2}{c} {$\bf i_{o} = 63.4^\circ$ }\\
  \includegraphics[width=6.0cm,height=4.75cm]{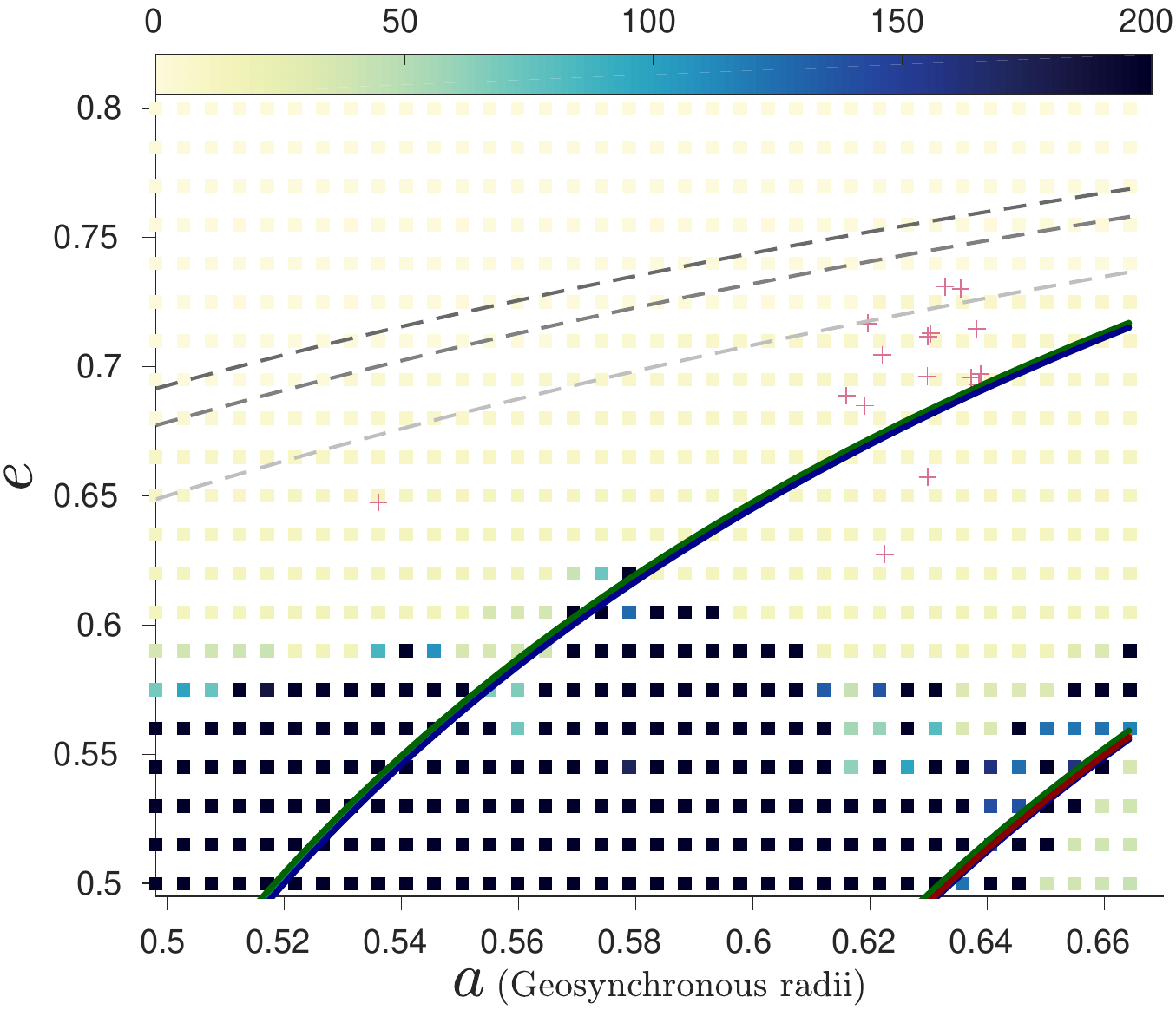} &
  \includegraphics[width=6.0cm,height=4.75cm]{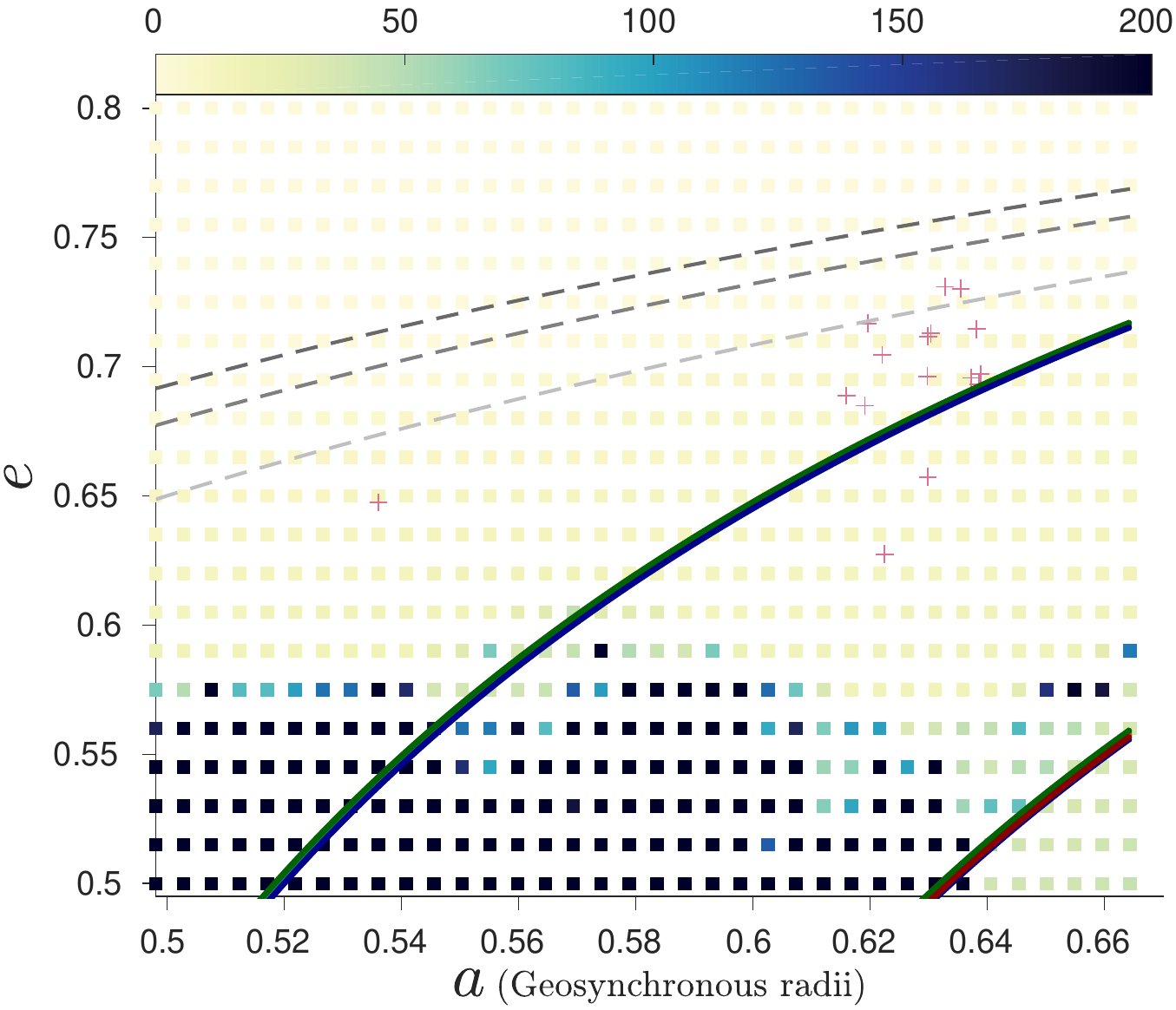}\\
\end{tabular}
\caption{Lifetime maps of the GTO region in $a-e$  for different initial $i$, and $\Delta\Omega = 180^\circ$, $\Delta\omega = 270^\circ$, Epoch 2020, $A/m = 
1$ m$^2$/kg, with atmospheric drag (right) or not (left).  The red points mark the cataloged population with $i=i_{o} \pm 5^{\circ}$ and $A/m\in\left(\left.0.02:1\right.\right]$ 
m$^2$/kg. The colorbar goes from lifetime 0 to 200 years.  The gray dashed curves correspond to $~100,~400,~1000\ km$ initial altitude ($q_0$, from high to low $e$). The multicolored bold lines correspond to the strongest lunisolar resonances that are present for each inclination. See text for more details.}
\label{fig:lifemaps_conf12_srp2}      
\end{figure}

We numerically integrated the trajectories of millions of orbits on centennial timescales, in order to investigate more thoroughly the long-term in the vicinity of GTOs. Our results are presented in the form of dynamical lifetime maps. A subset of these is shown in Figures \ref{fig:lifemaps_conf8_srp1}-\ref{fig:lifemaps_conf12_srp2}, where we compare the \textit{NO DRAG} and \textit{DRAG} cases, for specific inclinations and $A/m$ values. \\

Each map shows the grid of initial conditions in the $a-e$ plane for a given initial inclination, epoch, $A/m$ and phase angle  configuration. The colorbar indicates the dynamical lifetime of an orbit (i.e., the time spent before reaching our reentry limiting distance). On the same plot, red points represent the corresponding part of the cataloged population, as defined in Table~\ref{tab:objects}. The thin, dashed gray lines correspond to perigee altitudes $q_0=~100,~400,~1000$~km, going from higher to lower values of $e$, at given $a$. The multi-colored bold lines denote the locus of all relevant lunisolar resonances (up to 2nd order in the Legendre expansion) of the form $\dot{\psi}=j\dot{\omega}+k\dot{\Omega}+l\dot{\Omega}_{M}+mn_{S} = 0$,  that affect debris dynamics area near the $2:1$ tesseral resonance, where $\psi$ is the critical angle of the resonance, $\left(j,k,l,m\right)$ are integers, $\Omega_{M}$ is the node of the Moon, and $n_{S}$ the apparent mean motion of the Sun. Secular resonances in the near-Earth region have been mapped out and studied before in detail \citep[see][]{gC62,tEkH97,cC05,aCcG14,aCetal16,jD16,iGetal16}. A more complete study of the web of secular and semi-secular resonances (up to 2nd order) in the MEO region can be found in \citet{aR18}.\\
  
For low and moderate inclinations (i.e., Kourou, Cape Canaveral) and  low $A/m$, all particles with initial $q_0=a_0\left(1-e_0 \right)>R_{E}+100$~km, which is the limit that we set for reentry, remain `stable' in the \textit{NO DRAG} case, where by `stable' we mean that the eccentricity does not present any significant secular increase. The reentry region extends up to $q_0\approx R_{E}+500$~km in the \textit{DRAG} case, and the lifetime for the majority of reentries is less than $60$~yr. The results appear to be roughly independent of the initial orientation of secular angles, as there are no significant resonances to drive these objects to reentry conditions. Hence, from the corresponding part of the real population, only those with perigee altitude lower that $\sim 500$~km could naturally reenter within the next decades. The majority of objects with apogee $Q\approx 1\ a_{GEO}$, but with $q$ greater than the above limit, would remain stable. The results are similar when an augmented $A/m$ value is assumed. However, the reentry region extends to initial perigee altitudes of $500$~km and $1000$~km in the \textit{NO DRAG} and \textit{DRAG} cases, respectively. Also, the lifetimes of the majority of reentry particles goes down to $\sim 40$~yr. Comparing these results with the corresponding part of the current population, we can assert that most of these objects should reenter in the next few decades.\\

For high inclinations (i.e., Baikonur, Molniya), secular resonances play a greater role in the satellites' evolution, as overlapping can occur. These are the cases where the secular evolution of the orbits  show a strong dependence on the orientation phase angles. For $i\sim46^{\circ}$ and low $A/m$, the structure of the maps are similar in both the \textit{NO DRAG} and \textit{DRAG} cases. The value of initial perigee altitude that allows reentry can be larger than $1000$~km, and depends on the initial values of the secular angles. Reentry lifetimes have a large dispersion and may even extend beyond $100$~yr. When an augmented $A/m$ is considered, the initial perigee altitude that allows reentry can become larger than  $2000$~km and the lifetimes decrease.\\

For Molniya orbits, the results are particularly interesting, since two sets of overlapping resonances cross the corresponding $a-e$ projection for $i=63.4^{\circ}$.\footnote{Resonances are of the form $\dot{\psi}=j\dot{\omega}+k\dot{\Omega}+l\dot{\Omega}_{M} = 0$. Lines crossing through the point  $\left(a,e\right)=\left(0.518,0.5\right)$ correspond to $\left[j,k,l\right]=\left[0,1,-2\right],\left[2,1,-2\right],\left[2,-1,2\right]$. Lines passing through $\left(a,e\right)=\left(0.632,0.5\right)$ correspond to $\left[j,k,l\right]=\left[0,1,-1\right],\left[2,2,-2\right],\left[2,1,-1\right],\left[2,-1,1\right],\left[2,-2,2\right]$} The results are similar for both the \textit{NO DRAG} and \textit{DRAG} cases, and strongly depend on the orientation angles. In Figures~\ref{fig:lifemaps_conf8_srp1} and \ref{fig:lifemaps_conf12_srp1}, two different orientation configurations are shown. For $\Delta\Omega = 90^\circ$, $\Delta\omega = 270^\circ$, the population in the vicinity of the $2:1$ tesseral resonance with 
$e<0.68$ remains stable. On the other hand, for $\Delta\Omega = 180^\circ$, $\Delta\omega = 270^\circ$, the escape region is more extended and reentry can be achieved for orbits with initial $e>0.6$, on timescales of $\sim 50$~yr. For $A/m=1$~m$^2$/kg, the reentry region becomes even wider, though the map of lifetimes preserves its overall structure.  \\

\subsection{Examples of evolution}
\label{subsec:32}

\begin{figure}[htp!]
	\captionsetup{justification=justified}
	\centering 
 \begin{tabular}{cc}
  \includegraphics[width=0.4\textwidth]{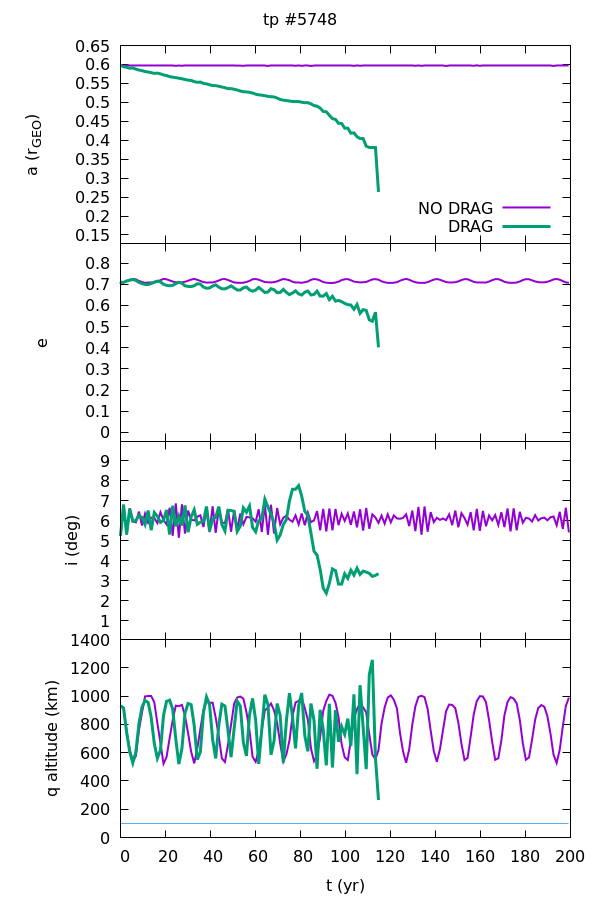} &
  \includegraphics[width=0.4\textwidth]{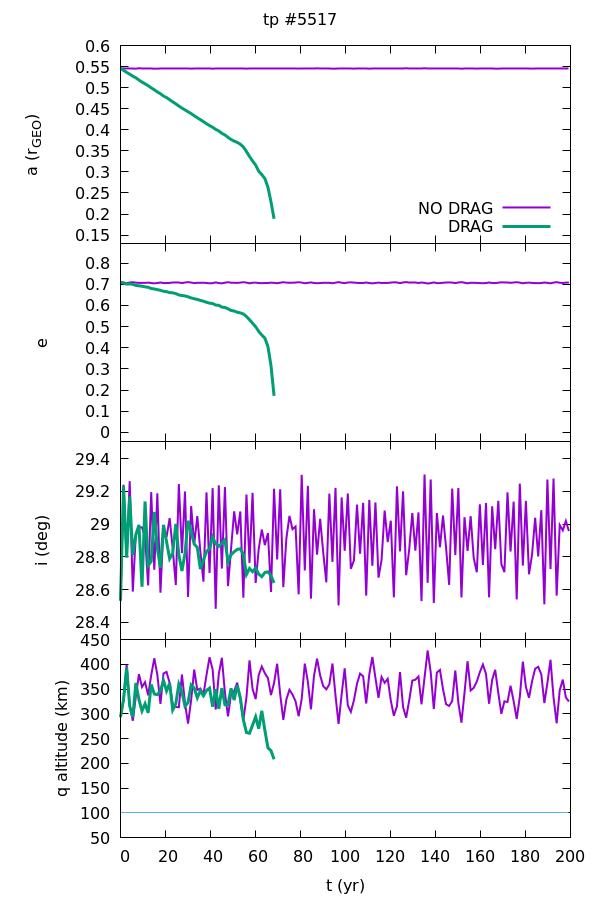} \\
  \includegraphics[width=0.4\textwidth]{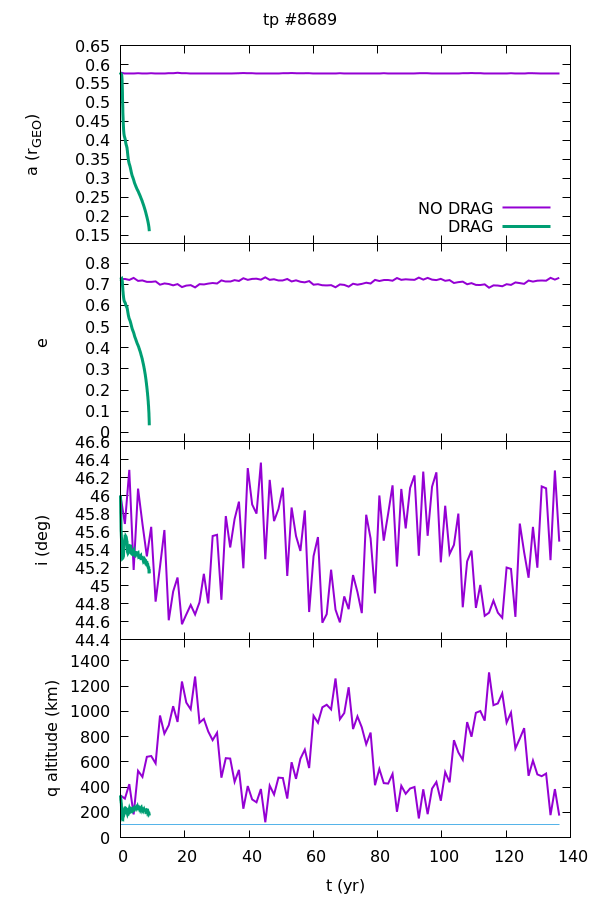} &
  \includegraphics[width=0.4\textwidth]{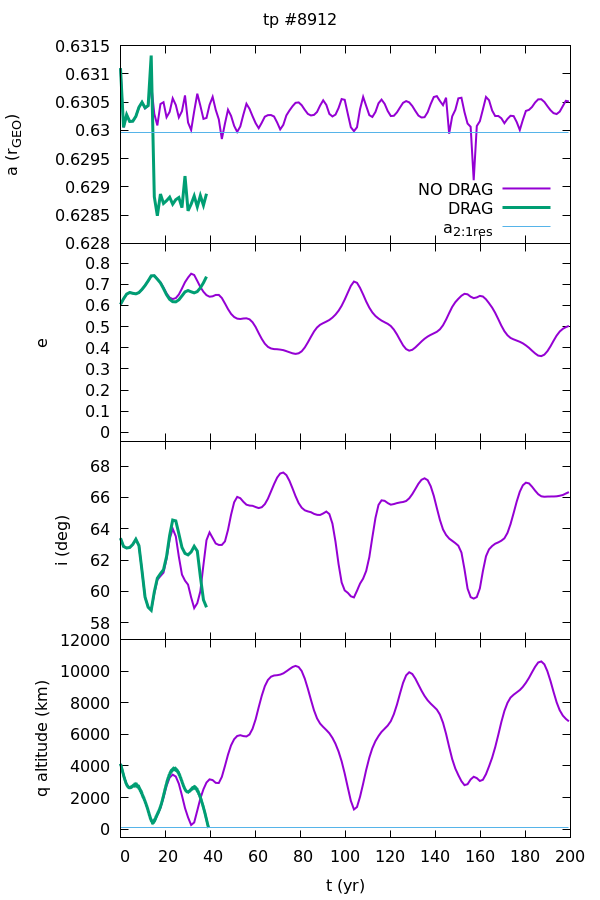} \\
 \end{tabular}
 \caption{ Evolution of semi-major axis, $a$, eccentricity, $e$, inclination, $i$, and apogee altitude, $q$, for particles with $i_0=5.235^\circ$ (top left; tp \#5748 with $a=0.598~a_{GEO}$, $e=0.710$, $\Omega=206.64^\circ$, $\omega=271.62^\circ$, $M=0$, and \textit{Epoch 2018}), $i_{0} = 28.533^\circ$ (top right; tp \#5517 with $a=0.546~a_{GEO}$, $e=0.710$, $\Omega=206.64^\circ$, $\omega=271.62^\circ$, $M=0$, and \textit{Epoch 2018}), $i_{0}=46^\circ$ (bottom left; tp \#8689 with $a=0.579~a_{GEO}$, $e=0.725$, $\Omega=268.87^\circ$, $\omega=7.47^\circ$, $M=0$, and \textit{Epoch 2020}) and $i_{0} = 63.4^\circ$ (bottom right; tp \#8912 with $a=0.631~a_{GEO}$, $e=0.725$, $\Omega=268.87^\circ$, $\omega=7.47^\circ$, $M=0$, and \textit{Epoch 2020}). For each particle, we compare the \textit{NO DRAG} (purple) and  \textit{DRAG} (green) cases. See text for details.}
 \label{fig:tp_evol1}      
\end{figure}

We present here some examples of orbital evolution that highlight the different dynamical mechanisms leading to reentry. Figure~\ref{fig:tp_evol1} shows the evolution of semi-major axis, $a$, eccentricity, $e$, inclination, $i$, and perigee altitude, $q$, of four particles with $i_{0}=5.235^\circ$ (top left), $i_{0} = 28.533^\circ$ (top right), $i_{0}=46^\circ$ (bottom left) and $i_{0} = 63.4^\circ$ (bottom right). All particles have $A/m=0.02$ m$^2$/kg, except the first one (i.e., $i_{o}=5.235^\circ$) which has $A/m=1$ m$^2$/kg. Aside from the `Baikonur' case shown here, all particles reenter only in presence of drag.\\

When drag is not included in the dynamical model, $a$ and $e$ evolve smoothly, except for the `Molniya' case. Drag significantly lowers $a$ and $e$ in the course of time and forces the orbit to reenter in the Earth's lower atmosphere. For the `Baikonur' GTO shown here, atmospheric drag speeds up its reentry by an order of magnitude. For the `Molniya' particle, the evolution during the first $20$~yr is similar in both dynamical models, but atmospheric drag forces the orbit to jump across the 2:1 tesseral resonance, to lower $a$ values. This, together with a favorable secular orientation leads to a substantial decrease in $q$, which then leads to an increased drag effect that eventually forces the orbit to reenter.

\subsection{Statistics}
\label{subsec:33}

From a first look -- and this is also supported by the figures shown so far -- dynamical lifetime values appear to have a large dispersion, while no clear correlation with initial conditions or other  parameters can be noticed. In previous studies (see e.g., \cite{vM12}) it was shown that dynamical lifetimes in the GTO region can vary by decades even for small variations, e.g.,\ in $A/m$ -- this strongly limits the usefulness of individual lifetime estimates. In our results, the most obvious correlation is with initial perigee altitude, as curves of (nearly) constant lifetime appear to smoothly follow $q=$const lines in our maps. However, as this behavior can vary significantly with inclination, $A/m$, epoch, and secular orientation, we define here some average (statistical) quantities that will enable us to get a deeper understanding of the mean behavior.\\
 
\begin{figure}[htp!]
	\captionsetup{justification=justified}
	\centering 
 \begin{tabular}{cc}
  \includegraphics[width=0.4\textwidth]{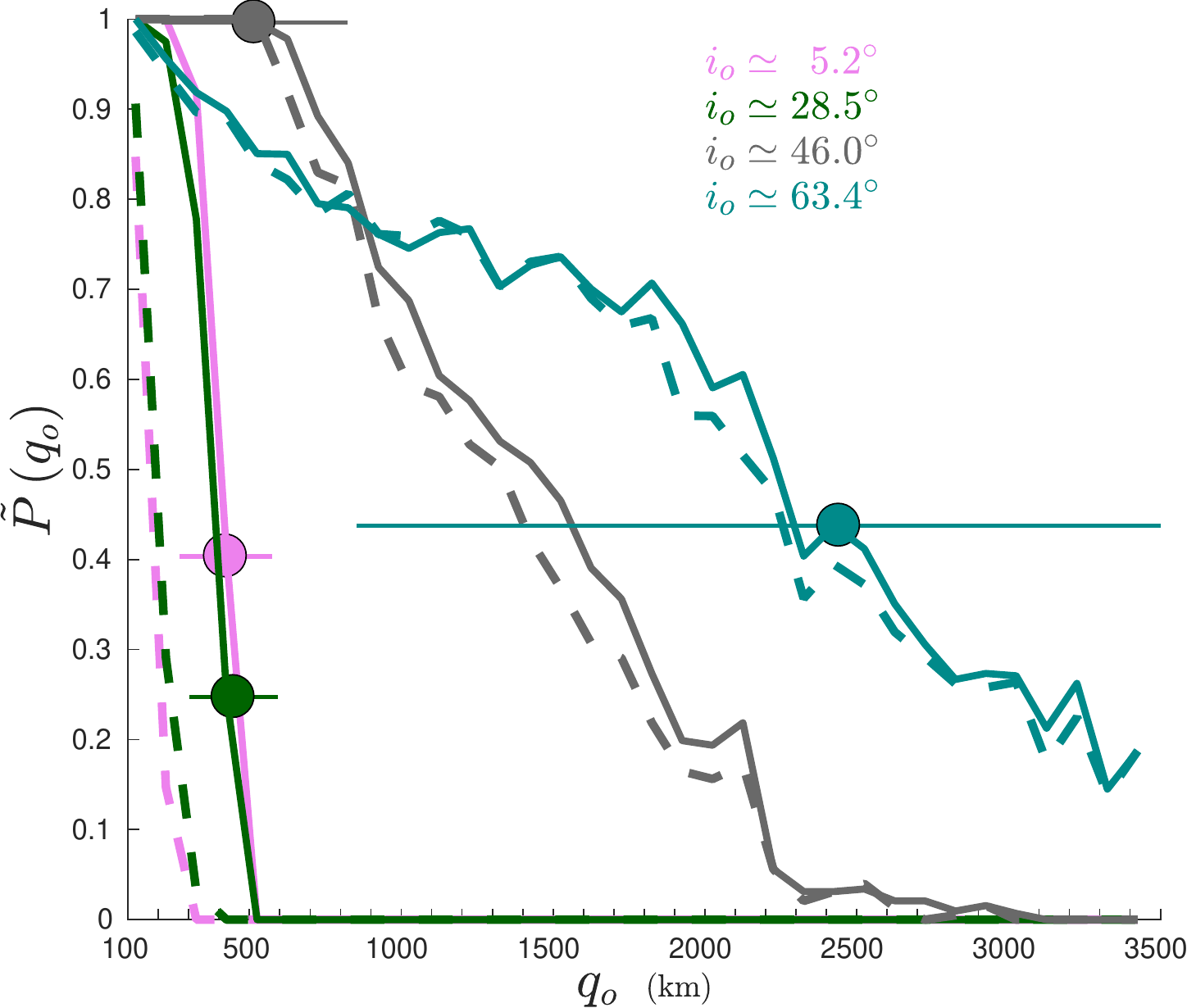} &
  \includegraphics[width=0.4\textwidth]{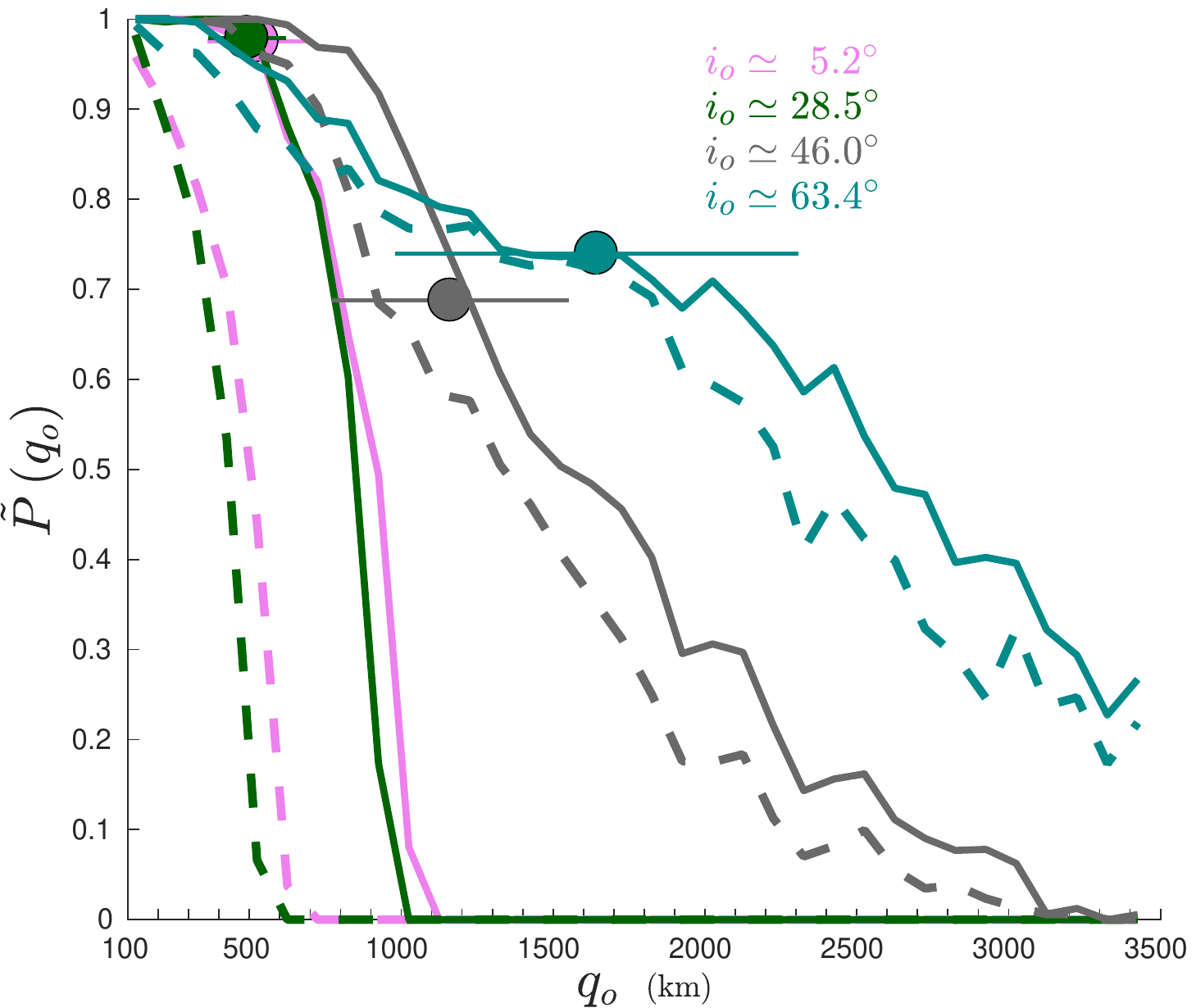} \\
 \end{tabular}
 \caption{Probability of reentry estimate, $\tilde{P}$, as function of initial perigee altitude (see text for details in calculating $\tilde{P}$).  (Left) $A/m = 0.02$ m$^2$/kg and (Right) $A/m = 1$ m$^2$/kg. Different colors denote different initial inclination; $i_{0}\simeq5.2^\circ$ (magenta), $i_{0}\simeq28.5^\circ$ (green), $i_{0}\simeq46.0^\circ$ (gray) and $i_{0}\simeq63.4^\circ$ (cyan). The dashed lines refer to the \textit{NO DRAG} case, whereas the solid ones refer to the \textit{DRAG} case. Solid circles placed on the \textit{DRAG} case's lines correspond to the mean value of perigee altitude of the corresponding part of the cataloged population; lines through denote the standard deviation.}
 \label{fig:propability_distr}      
\end{figure}

\begin{figure}[htp!]
	\captionsetup{justification=justified}
	\centering 
 \begin{tabular}{cc}
  \includegraphics[width=0.4\textwidth]{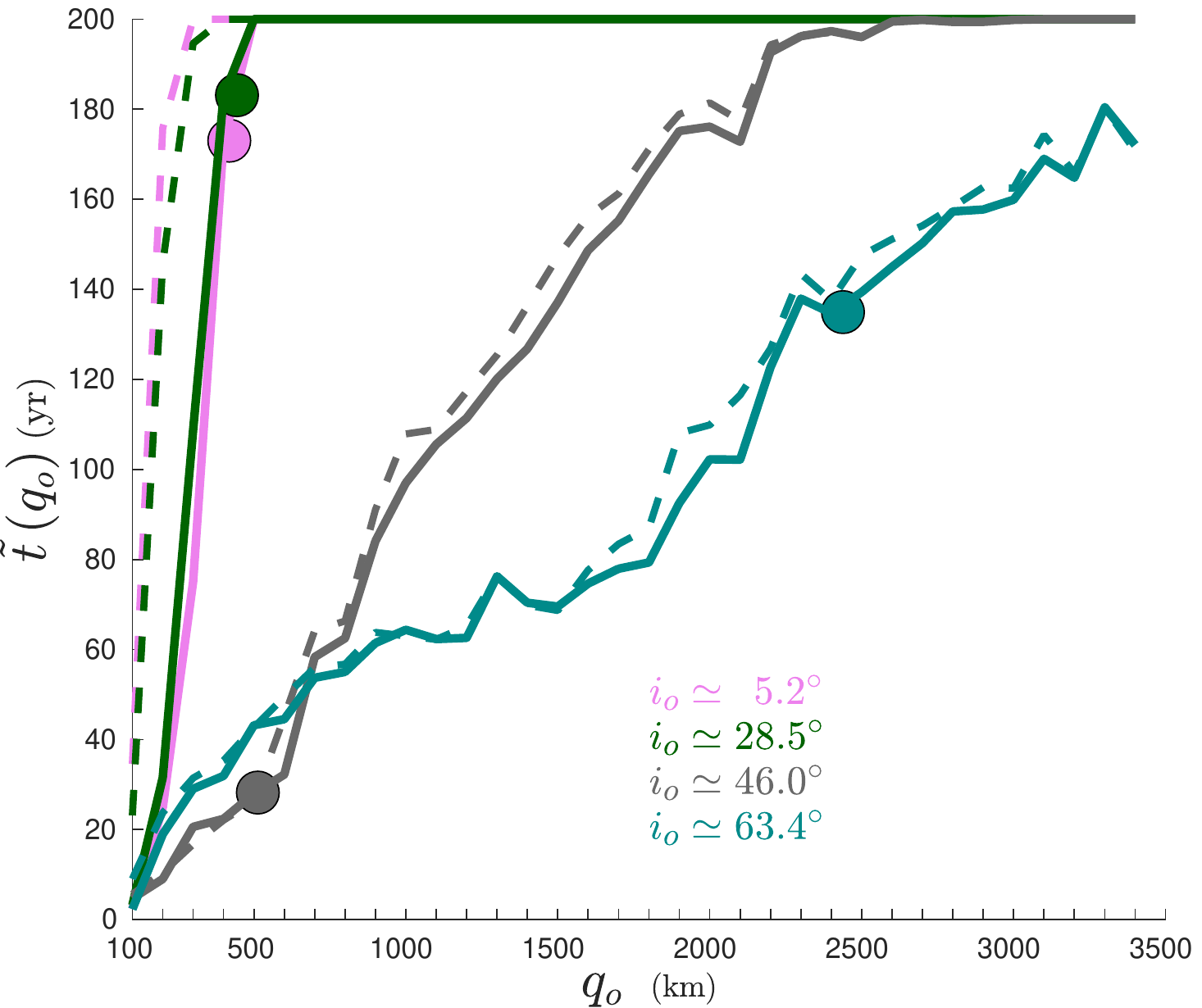} &
  \includegraphics[width=0.4\textwidth]{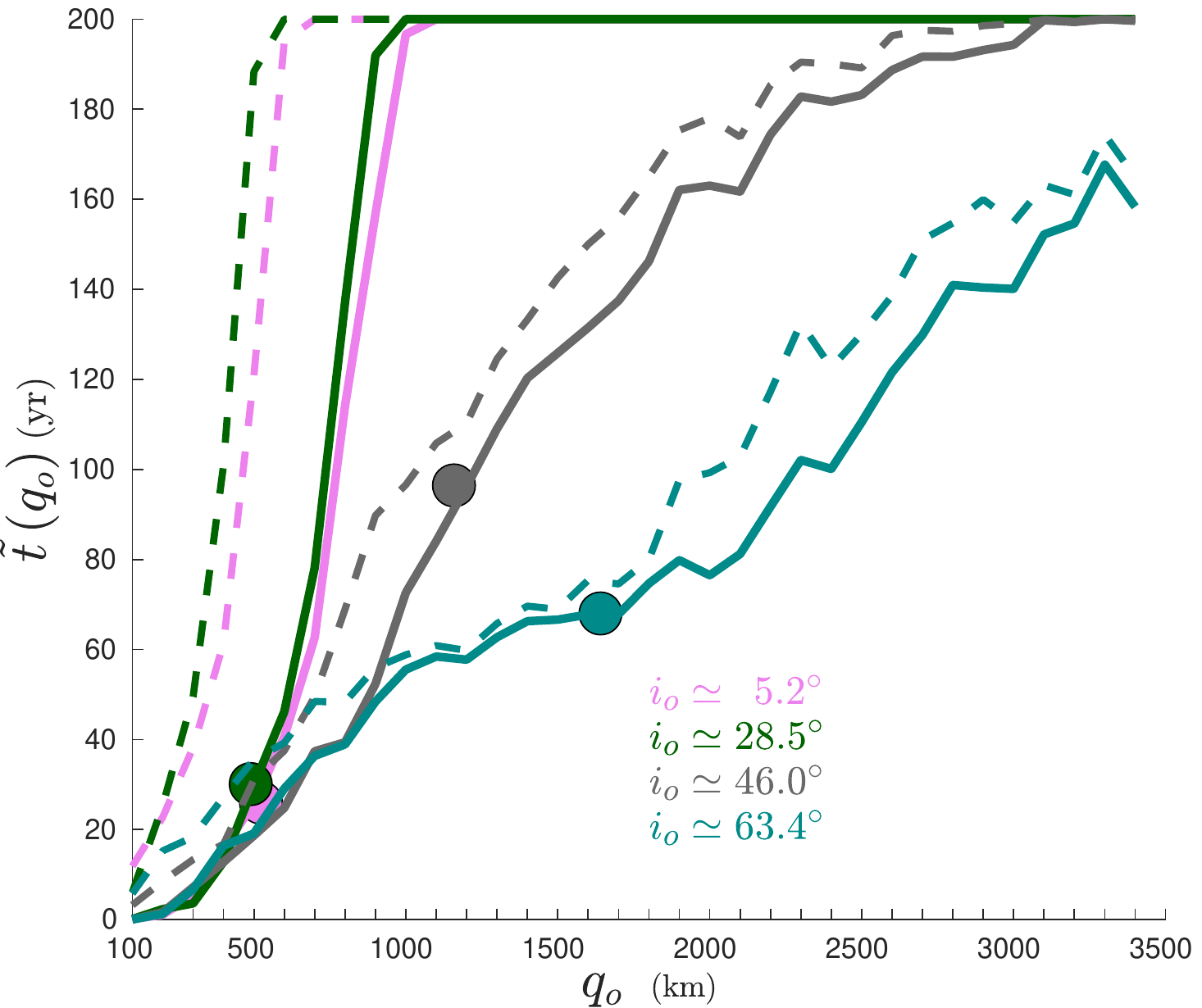} \\
 \end{tabular}
 \caption{Mean dynamical lifetime $\tilde{t}$ as function of perigee altitude (see text for details on its calculation). The color and line scheme is the same as in Figure \ref{fig:propability_distr}.}
 \label{fig:tlife_distr}      
\end{figure}

We compute the number of orbits that reenter for each $i_0$ and $A/m$ by summing up the results for all 16 secular configurations and for both epochs. We then split this number in perigee ($q_0$) bins of size $100~$km. Normalizing to the total number of orbits integrated in each bin, we derive an estimate of the probability density for reentry, as function of 
$q_0$ , $\tilde{P}(q_0)$. Similarly, we compute a lower bound for the mean dynamical lifetime, $\tilde t (q_0)$, by summing up the lifetimes of all particles in each bin and assuming a lifetime of $200$~years for particles that do not reenter, before normalizing to the total number of particles in each bin. Figure~\ref{fig:propability_distr} shows $\tilde{P}(q_0)$ and Figure~\ref{fig:tlife_distr} shows $\tilde{t}(q_0)$. Each curve in these figures corresponds to a different value of $i_0$, $A/m$, and dynamical model. Filled circles placed on the top of the \textit{DRAG} case lines correspond to the mean value of $q_0$ for the corresponding part of the cataloged population\footnote{For $i_{0}=5.2^\circ$ and $i_{0} = 28.5^\circ$, we only take into account the population with $q_{0}\le 800$~km, since, according to the dynamical maps presented in section \ref{subsec:31}, very few reentry solutions exist for higher altitudes.}. On the $\tilde{P}$ diagrams, the standard deviation of $q_0$ for the real objects is 
also denoted, by horizontal lines. \\

Let us summarize the basic characteristics seen in Figs.\ (11)-(12). Both $\tilde{P}$ and $\tilde{t}$ depend strongly on $q_0$. For $i_0\simeq5.2^{\circ}$ and $28.5^{\circ}$, both curves represent steep functions of $q_0$ but, for higher inclinations, they become shallower. Also, as the inclination becomes larger, the difference between the \textit{DRAG} and \textit{NO DRAG} cases becomes smaller. Finally, both functions depend on $A/m$ in two ways: they become shallower but, more importantly, they cross the $q_0$ axis at higher $q_0$ values, as $A/m$ increases -- i.e.,\ the reentry zone extends to higher altitudes. Note that this effect is also diminished for the two high-inclination groups. This indicates that the basic mechanism that controls reentry at high inclinations is the overlapping of lunisolar resonances.\\ 

Going into more details, at Kourou and KSC inclinations and for low $A/m$, the probability of reentry would go to zero at $q_0\gtrsim 300~$km, in the absence of drag. If drag is taken into account, the limiting value of perigee altitude becomes $q_0\approx 500~$km. An increase of $A/m$ by a factor of $50$ shifts these values to $\approx 500~$km (only by solar radiation pressure) and $\approx 1,000~$km, respectively. As a result, we estimate the reentry probability for debris with high $A/m$ (and $q_0< 1,000~$km) to $\tilde{P}\approx 1$, and with a mean lifetime $\tilde{t} \sim 25-30~$years. Note that, in all cases studied, the cumulative time spent by reentry orbits in the LEO protected area is very small ($\sim 1$~year). For low $A/m$ values (bigger objects), however, the reentry probability decreases ($\tilde{P}\approx 0.25-0.4$) and the mean lifetime increases significantly ($\tilde{t}\sim 170~$years). This only highlights the problem of mitigating large GTO objects, which is not significantly boosted by the natural long-term dynamics. \\

For high inclinations, $\tilde{P}$ has a much shallower dependence on $q_0$ and drops below $0.5$ at $\sim 1,500~$km and $2,000~$km for $i_0\simeq46.0^{\circ}$ and $i_0\simeq63.4^{\circ}$, respectively. As written above, these numbers do not change significantly when drag is considered and not even by increasing $A/m$ by a factor of $50$. Looking at the population of cataloged objects, we see that $\tilde{P}\approx 1$ for low-$A/m$ and $\approx 0.7$ for high-$A/m$ `Baikonur' objects, with dynamical lifetimes $\sim30$ and $\sim 90~$years, respectively. Note however that only 3 high-$A/m$ objects are registered in the catalog, at these inclinations\footnote{One might be tempted to interpret these few particles as long-lived remnants of an initially larger population}. These results suggests that Baikonur has a favorable latitude, in terms of large GTO mitigation, with respect to other launch sites. For Molniya-like objects with low $A/m$, the probability of reentry is $\sim 40$\% (but with a huge dispersion) and the mean lifetime is $\sim 130$~years. Increasing $A/m$, these numbers become $\sim75$\% and $\sim60$~years, respectively. These results suggest that small debris can effectively reenter from the Molniya region within a reasonable time-scale. The reentry process is also significant for larger Molniya objects, although less efficient by a factor of $\sim 2$, both in $\tilde{P}$ and $\tilde{t}$. \\

\section{Conclusions}
\label{concl}
We presented in this paper the results of an extensive numerical survey of dynamical lifetimes in the GTO phase-space region around the Earth;  this investigation is part of the ``ReDSHIFT'' project. The primary goal of our study was to record (maps of) the dynamical lifetime of these highly-elliptical orbits and its dependence on initial conditions and physical parameters, placing in context the cataloged population of that part of circumterrestrial space. \\

Our results show that debris with relatively high values of $A/m$ can naturally reenter within $20-100~$years with a probability of $>70\%$, practically from every inclination band; at low inclinations this applies only to objects with perigee altitude smaller than 1,000~km. Note that the probability estimation is based on the huge number of runs that we  carry out, by computing the ratio between favorable cases over the total number of cases. These numbers become much worse when typical, large (low $A/m$), low-inclination objects are considered. This is clearly the most problematic category of high-eccentricity, medium-Earth orbits, in terms of mitigation. On the other hand, Baikonur-launched GTOs seem to be the most efficient, in terms of self-mitigation. \\

Using our data-set of reentry solutions, we have computed `angle-averaged' estimates of the reentry probability and the mean dynamical lifetime for the different GTO `classes', as functions of initial perigee altitude, $q_0$, which we find to be the most critical parameter. At low-to-moderate inclinations, both quantities are quite steep, monotonically decreasing, functions of $q_0$. For higher inclinations, they become shallower, but still monotonically decreasing functions of $q_0$. Increasing $A/m$ and adding drag to the model basically `shift' the curves to higher $q_0$. However, the dependence on $A/m$ and drag is much smaller at high inclinations, where the main dynamical mechanism responsible for reentry is the overlapping of lunisolar resonances, as found in recent studies \citep[see][]{aCetal16,jD16,iGetal16}. \\

Our statistics depict quite well the mean behavior but can also help to interpret some features noted in previous studies, such as the strong variation of dynamical lifetime (by several decades) observed for small variations of $A/m$ (a few per cent) noted by \citep{vM12}. Our results suggest that this variation is much smaller at high inclinations. Moreover, at low inclinations, it is likely the result of the underlying steep dependence of lifetimes on $q_0$. Consider a given $\tilde{t} (q_0)$ curve for a given $i_0$, and a given $q_0=$const line that crosses the aforementioned curve at a high value of $\tilde{t}$, i.e.,\ at the `top' of the diagram. A small increase of $A/m$ would shift the $\tilde{t} (q_0)$ curve to the right; the lower $i_0$ the higher the shift. However, as the curves are very steep for low $i_0$ and the derivative with $q_0$ almost diverges, even a small shift of the curve would result in a huge variation of $\tilde{t}$, for a given initial condition; essentially our $q_0=$const line would end-up crossing the $\tilde{t}$ curve near the `bottom' of the diagram, rather than the `top'. \\

We did not present here a full treatment of the long-term dynamics of the extended GTO region, as our analysis on the possible dynamical mechanisms that lead to reentry is still ongoing. Also, more inclination bands and more values of $A/m$ and $C_D$ have to be studied, in 
order to fully understand the dependence of these mechanisms on these basic dynamical parameters and their efficiency in providing natural reentry solutions. We hope to report on this in a future publication.

\begin{acknowledgements} 
This research was funded by the European Commissions Horizon 2020, Framework Program for Research and Innovation (2014-2020), under Grant Agreement 687500 (project ReDSHIFT; http://redshift-h2020.eu/). The work of DKS is funded by the General Secretariat for Research and Technology (GSRT) and the Hellenic Foundation for Research and Innovation (HFRI). Results presented in this work have been produced using the AUTH Compute Infrastructure and Resources and the authors would like to acknowledge the support provided by the Scientific Computing Office throughout the progress of this research work. DKS wishes to acknowledge Davide Amato for supplying a prototype routine for atmospheric drag and for providing independent validation of our code.

\end{acknowledgements}


\end{document}